\documentclass[%
 reprint,
superscriptaddress,
 amsmath,amssymb,
 aps,
]{revtex4-2}

\usepackage{graphicx}
\usepackage{dcolumn}
\usepackage{bm}
\usepackage{siunitx}

\begin{document}

\preprint{APS/123-QED}

\title{Linking High-Harmonic Generation and Strong-Field Ionization in Bulk Crystals}
\author{Peter Jürgens}
\email{juergens@mbi-berlin.de}
\affiliation{Max-Born-Institute for Nonlinear Optics and Short Pulse Spectroscopy, Max-Born-Str. 2A, 12489 Berlin, Germany}
\affiliation{Advanced Research Center for Nanolithography (ARCNL), Science Park 106, 1098 XG Amsterdam, Netherlands}

\author{Sylvianne D.C. Roscam Abbing}
\affiliation{Advanced Research Center for Nanolithography (ARCNL), Science Park 106, 1098 XG Amsterdam, Netherlands}

\author{Mark Mero}
\affiliation{Max-Born-Institute for Nonlinear Optics and Short Pulse Spectroscopy, Max-Born-Str. 2A, 12489 Berlin, Germany}

\author{Graham G. Brown}
\affiliation{Max-Born-Institute for Nonlinear Optics and Short Pulse Spectroscopy, Max-Born-Str. 2A, 12489 Berlin, Germany}

\author{Marc J.J. Vrakking}
\affiliation{Max-Born-Institute for Nonlinear Optics and Short Pulse Spectroscopy, Max-Born-Str. 2A, 12489 Berlin, Germany}

\author{Alexandre Mermillod-Blondin}
\affiliation{Max-Born-Institute for Nonlinear Optics and Short Pulse Spectroscopy, Max-Born-Str. 2A, 12489 Berlin, Germany}

\author{Peter M. Kraus}
\affiliation{Advanced Research Center for Nanolithography (ARCNL), Science Park 106, 1098 XG Amsterdam, Netherlands}
\affiliation{Department of Physics and Astronomy, and LaserLaB, Vrije Universiteit, De Boelelaan 1105, 1081 HV Amsterdam, Netherlands}

\author{Anton Husakou}
\affiliation{Max-Born-Institute for Nonlinear Optics and Short Pulse Spectroscopy, Max-Born-Str. 2A, 12489 Berlin, Germany}

\date{\today}

\begin{abstract}
The generation of high-order harmonics in bulk solids subjected to intense ultrashort laser pulses has opened up new avenues for research in extreme nonlinear optics and light-matter interaction on sub-cycle timescales. Despite significant advancement over the past decade, a complete understanding of the involved phenomena is still lacking. High-harmonic generation in solids is currently understood as arising from nonlinear intraband currents, interband recollision and ionization-related phenomena. As all of these mechanisms involve or rely upon laser-driven excitation we combine measurements of the angular dependence of nonlinear absorption and high-order harmonic generation in bulk crystals to demonstrate the relation between high-harmonic emission and nonlinear, laser-induced ionization in solids.
 An unambiguous correlation between the emission of harmonics and laser-induced ionization is found experimentally, that is supported by numerical solutions of the semiconductor Bloch equations and calculations of orientation-dependent ionization rates using maximally localized Wannier-functions.
\end{abstract}

\maketitle


\section{Introduction}
High-order harmonic generation (HHG) in gases marked the birth of attosecond science allowing to track carrier dynamics on sub-cycle timescales \cite{Corkum_1993, Lewenstein_1994, Krausz_2009, Sansone_2010}. Transferring the concept of HHG to solid-state systems \cite{Ghimire_2011} has opened up several research avenues ranging from fundamental investigations of strong-field-driven carrier dynamics \cite{Schubert_2014} towards the development of compact extreme-ultraviolet (XUV) sources \cite{Garg_2018} and petahertz electronics \cite{Garg_2016, Ossiander_2022}. In solids, four main sources of nonlinearity have been identified and investigated with regards to harmonic generation. First, traditional perturbative nonlinearities \cite{Stolen_1973} have been invoked to explain conventional second harmonic generation (SHG) ever since Franken's seminal work \cite{Franken_1961} that marked the advent of nonlinear optics. It is based on the anharmonic motion of bound electrons in the valence band and is held responsible for low-order harmonic generation with multiple applications in frequency conversion as well as in optical-parametric chirped pulse amplification (OPCPA) \cite{Elu_2017}. Second, high-order harmonics extending to frequencies in the XUV \cite{Luu_2015} have been associated with interband recollisions (in strong analogy with the three-step model in the atomic case \cite{Krause_1992, Corkum_1993, Lewenstein_1994}). Third, high harmonics have also been explained by intraband currents where the nonlinearity enters through the non-parabolic shape of the conduction (and valence) bands \cite{Goulielmakis_2022}. Fourth, the nonlinearity that is inherent to the process of photoionization was proposed as a possible origin for harmonic generation \cite{Brunel_1990, Bauer_1998}. Three possible ionization-related mechanisms have been discussed in the context of HHG \cite{Geissler_1999, Juergens_2022}. Brunel harmonics, arising from the acceleration of quasi-free carriers excited by photoionization have been analyzed in gases, clusters, bulk solids and thin films \cite{Verhoef_2010, Babushkin_2022, Gao_2019, Mitrofanov_2011, Li_2022}. Moreover, after non-resonant photoionization an excited electron has nonzero velocity which provides sub-cycle contributions to the polarization and gives rise to the generation of harmonics. However, the strength of these velocity harmonics is expected to be insignificant under typical experimental conditions due to the modest photon energies that result in small excess velocities. Finally, the previously overlooked injection current, originating from the spatial displacement of an electron wavepacket during laser-driven tunneling ionization, was identified as the dominant source of low-order harmonic generation in fused silica \cite{Juergens_2020, Juergens_2022}.\\
The highly anisotropic angular dependence of the HHG process in crystalline solids has been correlated to the electronic band structure \cite{Luu_2015}, petahertz photocurrents \cite{Lanin_2019}, real-space trajectories of electronic wavepackets \cite{You_2017} and van-Hove singularities \cite{Uzan_2020, Suthar_2022}. Low-order (below-bandgap) harmonic emission has been associated with intraband dynamics while above-bandgap HHG is generally attributed to interband recollision \cite{Lanin_2019, Vampa_2015}. \\
All of the aforementioned HHG mechanisms - except for the Kerr-type nonlinearity - rely either on the presence or on the excitation of quasi-free conduction band electrons. In the most common experimental scheme the photon energy of the driving near-infrared or mid-infrared laser field $\hbar \omega$ is smaller than the bandgap $E_g$ of the irradiated crystal. Thus, excited carriers are generated by nonlinear photoionization or electron-impact ionization \cite{Vampa_2017}. Despite intense research in the past decade, the connection between carrier excitation and its impact on the anisotropic angular dependence of the harmonic emission has not yet been fully understood. While photoionization is often a key factor in harmonic generation, to date, there has been no investigation of its angular dependence under HHG conditions. \\
In this article we experimentally demonstrate a correlation between orientation-dependent ionization yields and the efficiency of HHG in various bulk crystals. We combine measurements of the angular dependence of nonlinear absorption (also referred to as multiphoton crystallography \cite{Gertsvolf_2008, Zhang_2021}) and angle-dependent HHG efficiencies to establish a direct link between the excitation of quasi-free carriers and the emission of high-order harmonics. We identify an unambiguous connection between intense HHG emission and strong ionization for distinct orientations in various oxide and fluoride crystals. Our experimental results are supported by numerical simulations based on the semiconductor Bloch equations (SBEs) and calculations of the photoionization rate based on the Wannier-formalism. Our numerical results reproduce the experimentally observed angular dependence of the harmonic emission and provide insights into the complex interplay of interband and intraband dynamics as well as the importance of ionization-related mechanisms for low-order harmonic generation.
\section{Experimental Setup}
In the experiments we used the signal beam of a home-built, high-repetition-rate, dual-beam OPCPA (\SI{100}{\kilo\hertz}) with a central wavelength of \SI{1500}{\nano\meter} and a full-width half-maximum (FWHM) pulse duration of $\sim$\SI{50}{\femto\second} (architecture based on the one presented in Ref.~\cite{Heiner_2018}) to generate high-order harmonics from bulk crystals. The linearly polarized beam was focused to an $1/{e^2}$ beam diameter of $\sim$\SI{50}{\micro\meter} (measured by the knife-edge technique in ambient air) providing a maximum peak intensity in ambient air of $\sim$\SI{30}{\tera\watt\per\centi\meter\squared} in the focal plane. 
\begin{figure}[ht]
\centering\includegraphics[width=\linewidth]{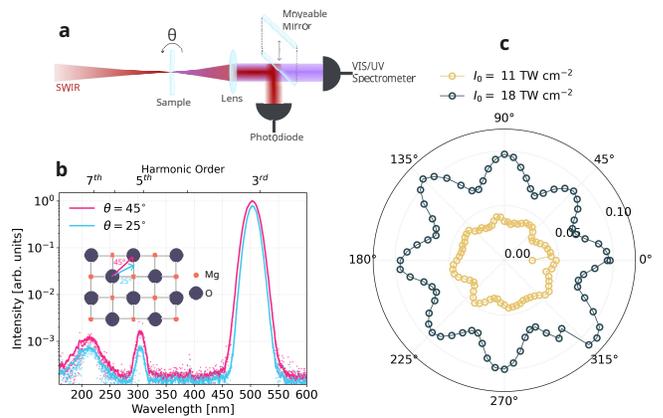}
\caption{Experimental setup and results obtained in a bulk MgO crystal. (a). A short-wavelength infrared (SWIR) laser pulse ($\lambda_0=\,$ \SI{1500}{\nano\meter}, $\tau=\,$\SI{50}{\femto\second}) is focused into the bulk of a crystal mounted on a rotation stage. The harmonic radiation is analyzed using a visible-ultraviolet (VIS/UV) spectrometer. The transmitted fundamental radiation can be analyzed by inserting a moveable dielectric mirror directing the SWIR beam onto a photodiode. (b). Illustrative harmonic spectra containing the third, the fifth and the seventh harmonic of the SWIR driving field obtained in a bulk MgO crystal at two different orientations ($\theta =\,$\SI{25}{\degree} and \SI{45}{\degree}) and a peak intensity of \SI{18}{\tera\watt\per\centi\meter\squared}. The large spectral width of the seventh harmonic is attributed to spectral broadening in the detection system. (c). Absorption of the fundamental laser field for different peak intensities below the damage threshold as a function of the orientation angle $\theta$. }
\label{fig:fig_1}
\end{figure}
High harmonics emitted from bulk crystals were analyzed in transmission geometry with the help of a commercial visible-ultraviolet (VIS-UV) spectrometer (Avantes AvaSpec-HS1024x58/122TEC). The energy remaining in the short-wave infrared (SWIR) signal beam was directed onto a photodiode [cf. Fig.~\ref{fig:fig_1}(a)] with the help of a highly reflective mirror at \SI{1500}{\nano\meter}. A possible contamination of the transmitted signal beam by harmonic radiation was excluded due to the low reflectivity of the dielectric mirror for the harmonic wavelengths and by the low conversion efficiencies (typically around $10^{-6}$\cite{Gholam_2017}). The crystalline samples were mounted on a computer-controlled rotation stage enabling a precise and reproducible variation of the angle $\theta$ between the main axis of the crystal and the laser polarization axis.  
\section{Experimental Results}
\begin{figure}[ht]
\centering\includegraphics[width=\linewidth]{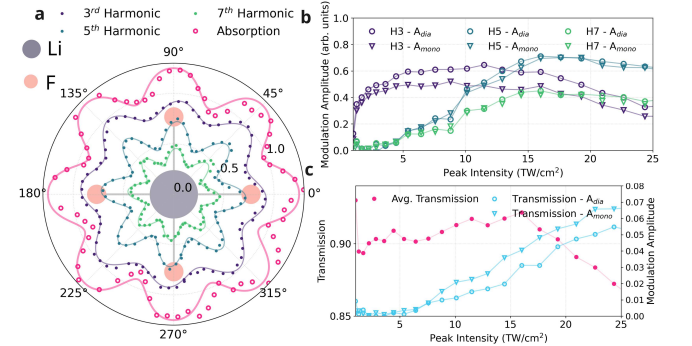}
\caption{Experimentally determined HHG and absorption in LiF. (a) $\theta$-dependence of HHG and SWIR absorption in LiF. (b) Fitted amplitudes $A_{\rm{mono}}$ and $A_{\rm{dia}}$ as a function of the SWIR intensity. (c) Average transmission and modulation amplitude of the 8-fold symmetric transmission loss.}
\label{fig:fig_4}
\end{figure}

Figure~\ref{fig:fig_1}(b) and (c) show exemplary high-harmonic spectra obtained in a \SI{200}{\micro\meter} thick MgO crystal, cut along the (100) plane, at two different orientation angles ($\theta = \,$\SI{25}{\degree} and \SI{45}{\degree}, for minimum and maximum signal strength) and the absorption of the SWIR laser pulse as a function of $\theta$, for two different peak intensities ($I_1 = \,$ \SI{11}{\tera\watt\per\centi\meter\squared}, $I_2 = \,$ \SI{18}{\tera\watt\per\centi\meter\squared}). The harmonic yield of all observed orders is higher for $\theta = \,$\SI{45}{\degree} when compared to $\theta = \,$\SI{25}{\degree} (the crystal angles were calibrated by perturbative third harmonic generation at a moderate excitation intensity of $\sim$\SI{2}{\tera\watt\per\centi\meter\squared}). 
Based on the inset shown in Fig.~\ref{fig:fig_1}(b) illustrating the cubic crystal structure of MgO, such a yield anisotropy can be attributed to high-symmetry directions in the crystal lattice. At an angle of \SI{45}{\degree} the real space trajectories of the excited electron wavepackets point towards the nearest-neighbour of the same species (Mg-Mg or O-O) inside the crystal structure (monoatomic nearest-neighbour direction).\\
The SWIR absorption exhibits no pronounced $\theta$-dependence at a low peak intensities below $\sim\,$\SI{8}{\tera\watt\per\centi\meter\squared}. Instead, the absorption stays close to 0 after accounting for the transmission losses due to Fresnel reflection at the interfaces ($\sim\,$\SI{16}{\percent} for MgO). At an intermediate SWIR excitation intensity of \SI{11}{\tera\watt\per\centi\meter\squared} the average absorption increases to $\sim$\SI{4}{\percent} [yellow circles in Fig.~\ref{fig:fig_1}(c)]. At the same time distinct extrema begin to emerge. When the excitation intensity approaches the laser-induced damage threshold of the used samples (determined to be $\sim$\SI{20}{\tera\watt\per\centi\meter\squared}), the modulation depth of these oscillations increases and a clear eight-fold symmetry becomes apparent in the $\theta-$dependent absorption of the SWIR pump laser pulse with maxima appearing in the diatomic ($\theta = \,$\SIlist{0;90;180;270}{\degree}) and monoatomic ($\theta = \,$\SIlist{45;135;225;315}{\degree}) nearest-neighbour directions. \\
\begin{figure}[ht]
\centering\includegraphics[width=\linewidth]{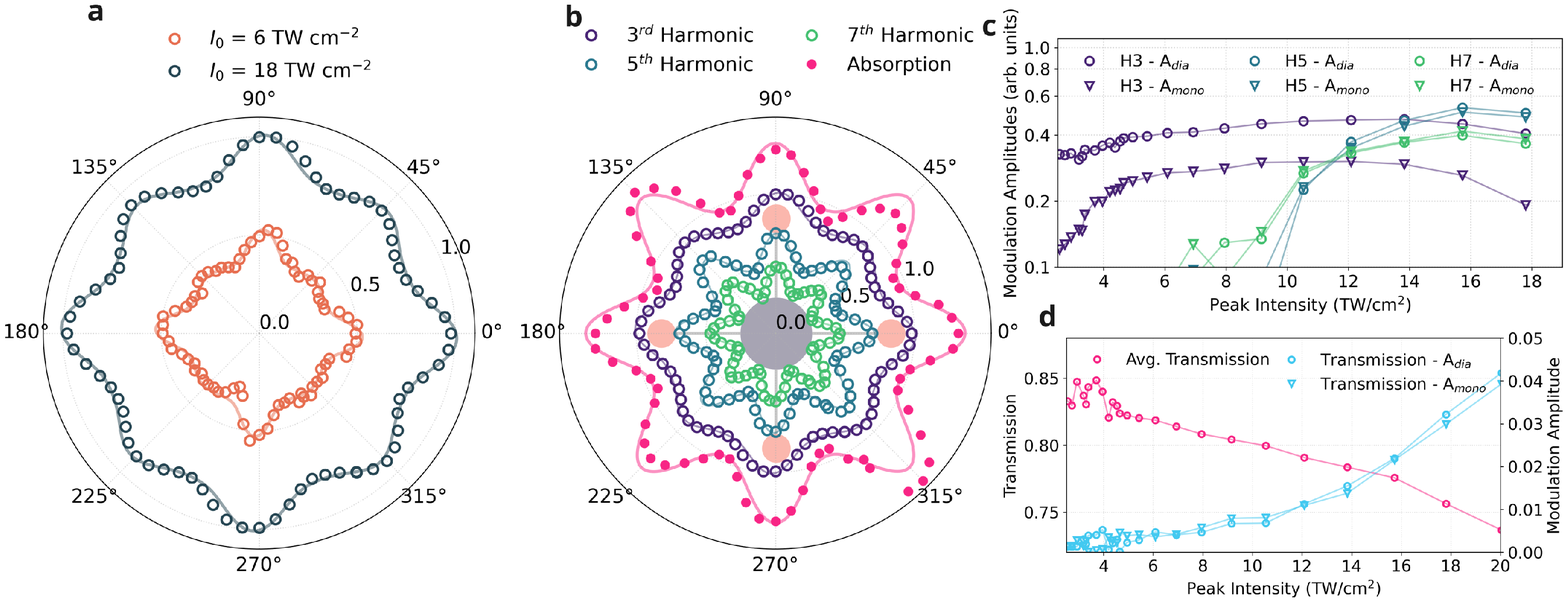}
\caption{Experimentally determined HHG and absorption in MgO. (a) Angular dependence of the third harmonic in the perturbative (at $I_0 =\,$ \SI{6}{\tera\watt\per\centi\meter\squared}, orange circles) and in the non-perturbative regime (at $I_0 =\,$ \SI{18}{\tera\watt\per\centi\meter\squared}, grey circles). (b) $\theta$-dependence of the observed odd harmonics and the SWIR absorption in a \SI{200}{\micro\meter} thick MgO crystal obtained at a peak intensity of \SI{18}{\tera\watt\per\centi\meter\squared}. The cubic crystal structure of MgO is sketched in the background to indicate the nearest-neighbour directions. The harmonic signals as well as the absorption are normalized and offset vertically for clarity. (c) Amplitudes extracted from the fit function (Eq.~\ref{eq:fit_function}) for the observed harmonic orders as a function of the SWIR excitation intensity. (d) Average transmission and modulation amplitude of the fundamental pump laser pulse as a function of intensity.}
\label{fig:fig_2}
\end{figure}
\\
With the goal to link the nonlinear absorption and the emission of high-order harmonics, the orientation-dependent harmonic yields are compared to the nonlinear absorption in Fig.~\ref{fig:fig_2}. \\
Similar to the angular distribution of the absorption, the third harmonic signal exhibits an eight-fold symmetry at a SWIR peak intensity of \SI{18}{\tera\watt\per\centi\meter\squared} [see dark green circles in Fig.~\ref{fig:fig_2}(a)]. As reported in Ref.~\cite{You_2017} the maxima of the third harmonic emission along the monoatomic nearest-neighbour directions ($\theta =\,$\SIlist{45;135;225;315}{\degree}) are suppressed at low intensity where perturbative mechanisms dominate the nonlinear response [orange circles in Fig.~\ref{fig:fig_2}(a)]. Hence, a transition from a four-fold symmetry to an eight-fold symmetry can be observed for the third harmonic as the excitation intensity increases. \\
At an excitation intensity of \SI{18}{\tera\watt\per\centi\meter\squared} the harmonic yield of all observed orders is maximized along the diatomic nearest-neighbour directions (Mg-O bonding directions at $\theta = \,$\SIlist{0,;90;180;270}{\degree}) as shown in Fig.~\ref{fig:fig_2}(b). Further maxima are formed along the monoatomic nearest-neighbour directions (at $\theta=$ \SI{45}{\degree}, \SI{135}{\degree}, \SI{225}{\degree} and \SI{315}{\degree}) resulting in an eight-fold symmetry of all observed harmonic orders whose extrema align exactly with those of the nonlinear absorption [pink circles in Fig.~\ref{fig:fig_2}(b)] suggesting that the harmonic emission is directly correlated with nonlinear photoionization. 
Due to the lower signal strength the fifth and seventh harmonic could only be observed at intensities where all orders exhibited an eight-fold symmetry. \\
For further analysis the angle-dependent harmonic yields $Y(\theta)$ as well as the $\theta$-dependent absorption are approximated by a periodic fit function according to
\begin{equation}
\label{eq:fit_function}
    Y(\theta) = A_0 + A_{\rm{dia}} * \cos[2 \theta]^{2n} + A_{\rm{mono}} * \sin[2\theta]^{2n} .
\end{equation}
Here $A_0$ is an offset amplitude, $A_{\rm{dia}}$ and $A_{\rm{mono}}$ denote the amplitudes associated with the oscillations along the mono- and diatomic nearest-neighbour directions, respectively. The parameter $n$ determines the width of the maxima and is well approximated by 3 throughout this work. \\
In Fig.~\ref{fig:fig_2}(c) the amplitudes that were introduced in Eq.~\ref{eq:fit_function} are analyzed as a function of the SWIR intensity. The eight-fold angular dependence of the fifth and the seventh harmonic contains equal contributions of mono- and diatomic amplitudes. However, for the third harmonic $A_{\rm{mono}}$ and $A_{\rm{dia}}$ evolve differently. While $A_{\rm{dia}}$ constantly exhibits values between 0.3 and 0.5 over the full range of intensities, $A_{\rm{mono}}$ sharply increases up to an SWIR intensity of $\sim\,$\SI{4.5}{\tera\watt\per\centi\meter\squared} before following the same trend as $A_{\rm{dia}}$ at a slightly ($\sim 0.15$) lower level. This intensity-dependent behavior is consistent with the aforementioned transition from a four-fold to an eight-fold symmetry as shown in Fig.~\ref{fig:fig_2}(a). \\
Figure~\ref{fig:fig_2}(d) demonstrates how the transmission of the SWIR pulse decreases as the excitation intensity increases towards the laser-induced damage threshold of the MgO crystal. At the same time, the modulation amplitude (for the absorption $A_{\rm{mono}} \approx A_{\rm{dia}}$) of the eight-fold symmetry increases with intensity in a manner that is notably different from the intensity dependence of the amplitudes that were discussed in Fig.~\ref{fig:fig_2}(c). This is consistent with the fact, that although the transmission is related to the frequency conversion process, these processes only account for a negligible part of the transmission loss. \\
\begin{figure}[ht]
\centering\includegraphics[width=\linewidth]{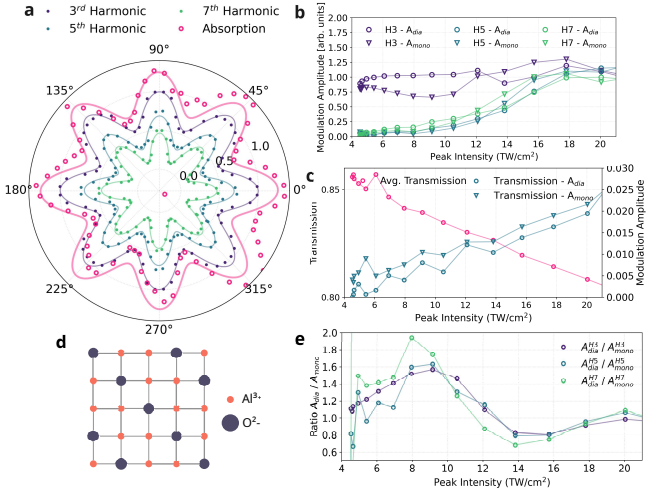}
\caption{Experimentally determined HHG and absorption in Al$_2$O$_3$. (a) $\theta$-dependence of HHG and SWIR absorption in Al$_2$O$_3$ at an excitation intensity of \SI{18}{\tera\watt\per\centi\meter\squared}. (b) Fitted amplitudes $A_{\rm{mono}}$ and $A_{\rm{dia}}$ as a function of the SWIR intensity. (c) Average transmission and modulation amplitude of the 8-fold symmetric transmission loss. (d) Sketch of the projection of the hexagonal unit cell of an M-cut Al$_2$O$_3$ sample. (e) Ratio of the mono- and diatomic amplitudes extracted from the fit function (Eq.~\ref{eq:fit_function}) showing the transition from a stronger $A_{\rm{dia}}$ to a stronger $A_{\rm{mono}}$ with increasing intensity, finally settling at $A_{\rm{dia}} \sim A_{\rm{mono}}$.}
\label{fig:fig_3}
\end{figure}
\\
We repeated the same experiments in bulk sapphire samples (\SI{150}{\micro\meter} thick) cut along the M-plane ($10\overline{1}0$) for which the projection of the hexagonal unit cell onto the M-plane exhibits a similar structure as MgO [see Fig.~\ref{fig:fig_3}(d)]. For an excitation intensity of \SI{18}{\tera\watt\per\centi\meter\squared} (damage threshold at $\sim\,$ \SI{21}{\tera\watt\per\centi\meter\squared}) we observe an eight-fold symmetry for all detected harmonic orders as well as for the nonlinear absorption of the fundamental SWIR laser pulse [see Fig.~\ref{fig:fig_3}(a)]. Again, we observe maxima in the angular distribution of the SWIR absorption when the high-harmonic emission is maximized. Even though the orientation angles of the HHG and absorption maxima do not correspond to mono- and diatomic nearest-neighbour directions we keep the terminology that was introduced in the MgO case. Figure~\ref{fig:fig_3}(b) shows the modulation amplitudes extracted from Eq.~\ref{eq:fit_function} as a function of the SWIR peak intensity. Even at the lowest intensity of $\sim\,$\SI{4}{\tera\watt\per\centi\meter\squared} the third harmonic signal exhibits an eight-fold symmetry resulting in comparable magnitudes of $A_{\rm{mono}}$ and $A_{\rm{dia}}$ [see blue circles and blue triangles in Fig.~\ref{fig:fig_3}(b)]. For all observed harmonic orders $A_{\rm{mono}}$ and $A_{\rm{dia}}$ display a similar intensity dependence. We find a transition from a stronger diatomic amplitude for intensities below \SI{12}{\tera\watt\per\centi\meter\squared} to a stronger monoatomic amplitude, before, at the highest measured intensities, the two amplitudes become equal again. This observation is confirmed by plotting the ratio $A_{\rm{dia}}/A_{\rm{mono}}$ in Fig.~\ref{fig:fig_3}(e). The ratio changes from values $\geq\,$1 for excitation intensities up to \SI{12}{\tera\watt\per\centi\meter\squared} to ratios $\leq\,$1 for intensities between \SI{12}{\tera\watt\per\centi\meter\squared} and \SI{18}{\tera\watt\per\centi\meter\squared}. This indicates that the emission from real-space trajectories towards monoatomic nearest-neighbours at $\theta = \,$\SIlist{45;135;225;315}{\degree} becomes more important than the emission from shorter diatomic nearest-neighbour trajectories at $\theta = \,$\SIlist{0;90;180;270}{\degree} for $I_0 \geq\;$\SI{12}{\tera\watt\per\centi\meter\squared}. The transmission results obtained from the sapphire sample display the same characteristics as in the MgO case with the exception that $A_{\rm{mono}}$ is constantly slightly larger than $A_{\rm{dia}}$ [see Fig.~\ref{fig:fig_3}(c)]. \\
In order to verify the generality of our experimental approach we furthermore analyzed the angular dependence of the HHG process and the nonlinear absorption in wide-bandgap fluoride crystals. As a prominent example we used LiF crystals (\SI{200}{\micro\meter} thickness) with a cubic  crystal structure similar to the one of MgO [see inset in Fig.~\ref{fig:fig_4}(a)]. Apart from the fact that no transition from a four-fold to an eight-fold symmetry was observed for the third harmonic, the results shown in Fig.~\ref{fig:fig_4} exhibit comparable characteristic features as in MgO. Both, the observed odd harmonic orders as well as the nonlinear absorption display an eight-fold symmetry in which the maxima of the HHG signal coincide with those of the SWIR absorption [Fig.~\ref{fig:fig_4}(a)]. The angular contrast of the odd harmonics and of the absorption reaches comparable values as in the MgO case [see Fig.~\ref{fig:fig_4}(b)] and the average transmission shows the characteristic drop from the Fresnel-related losses at low intensity to lower values at higher intensity towards the damage threshold [see Fig.~\ref{fig:fig_4}(c)]. \\
The experimental results presented in this section provide consistent evidence for a correlation between the emission of high-order harmonics and nonlinear absorption of the SWIR pump laser pulse. In all investigated materials the harmonic emission was the strongest at orientation angles where also the nonlinear absorption was at its peak. We interpret these findings as an indication that ionization, which results from the nonlinear absorption, drives the maximization of the HHG process. In MgO a transition from a four-fold to an eight-fold symmetry was observed for the angular dependence of the third harmonic as the SWIR intensity increased. In Al$_2$O$_3$ all detected harmonic orders exhibited an eight-fold symmetry irrespective of the pump laser intensity. However, a change of the strongest emission angles from $\theta=\,$\SIlist{0;90;180;270}{\degree} (called diatomic nearest-neighbour directions in the MgO case) to $\theta=\,$\SIlist{45;135;225;315}{\degree} (monoatomic nearest-neighbour directions) was identified. \\

Based on our experimental results we identified an unambiguous link between HHG and laser-induced nonlinear ionization. However, all widely discussed mechanisms for HHG in solids (interband recollision, intraband dynamics and ionization-related mechanisms) require the presence of quasi-free, excited electrons in the conduction bands. Hence, an unambiguous identification of the dominating HHG mechanism solely based on the presented experimental results is out of reach. \\
\section{Numerical Results and Discussion}
\begin{figure}[ht]
\centering\includegraphics[width=\linewidth]{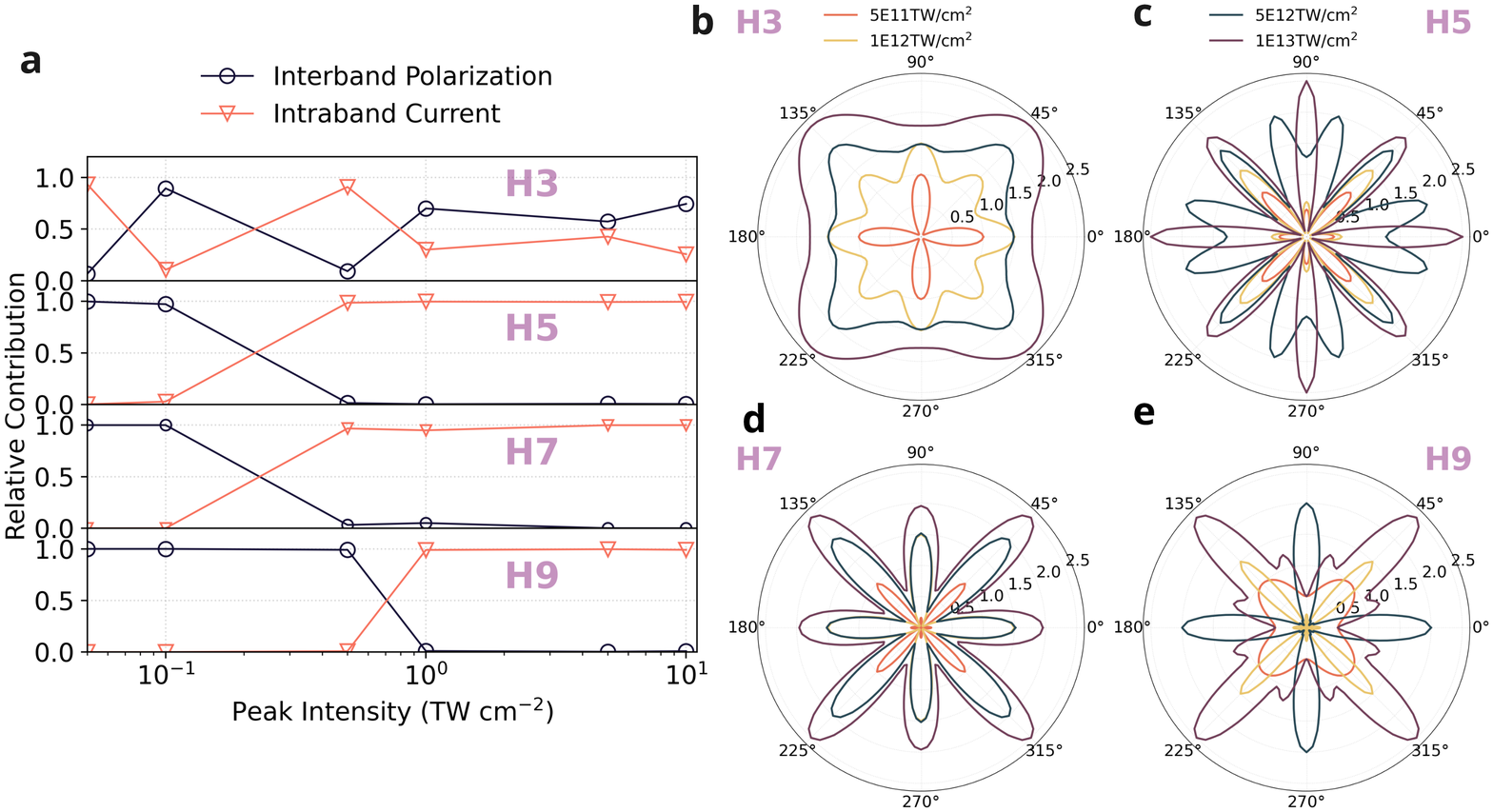}
\caption{Numerical simulations of HHG in MgO using SBEs. (a) Relative importance of the interband polarization and the intraband current for the total HHG signal obtained by integrating the harmonic yields for $\theta \in [0,2 \pi]$. (b) - (e) Orientation-dependence of the odd harmonics (H3-H9) for various SWIR peak intensities.}
\label{fig:fig_5}
\end{figure}

\begin{figure}[ht]
\centering\includegraphics[width=\linewidth]{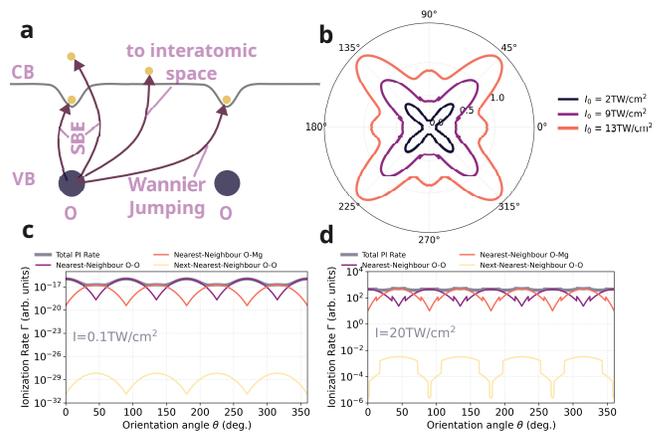}
\caption{Numerical simulations of HHG in MgO using maximally-localized Wannier-functions. (a) Depiction of electronic excitation at a single atom (as captured by the SBEs), to interatomic space and to neighbouring atoms (Wannier-Jumping). (b) Orientation-dependence of the photoionization rate calculated according to Eq.~\ref{eq:mpi_rate} for different SWIR laser intensities. (c) \% (d) Angular dependence of the photoionization rate of the various nearest-neighbour and next-nearest neighbour transitions at a SWIR peak intensity of \SI{0.1}{\tera\watt\per\centi\meter\squared} (c) and \SI{20}{\tera\watt\per\centi\meter\squared} (d).}
\label{fig:fig_6}
\end{figure}

To improve our understanding of the dominant HHG mechanism and of the observed $\theta$-dependent features we performed numerical simulations based on the SBEs \cite{Lindberg1988, Yu_2016} for the case of MgO. In our simulations we implement the band structure of MgO (taken from Ref.~\cite{You_2017}, using one valence and one conduction band) to estimate the $\theta$-dependence of HHG by projecting the electric field of the SWIR pump laser pulse onto the $\Gamma-K$ and $\Gamma-X$ directions. We then calculate the resulting high-harmonic emission due to the interband polarization $P(\omega)$ and the intraband current $J(\omega)$ (details of the numerical model can be found in Ref.~\cite{Abbing_2022}). The relative contribution of the two competing mechanisms is displayed in Fig.~\ref{fig:fig_5}(a). Strikingly, the intraband current (orange triangles) dominates the emission of H5, H7 and H9 for SWIR peak intensities $\geq\,$\SI{1}{\tera\watt\per\centi\meter\squared} while for the third harmonic the emission due to the interband polarization (dark circles) remains stronger. Thus, contrary to previous publications \cite{Lanin_2019, Schubert_2014, Vampa_2015, Vampa_2015_a, Goulielmakis_2022}  we do not observe a dominant intraband mechanism for all observed below-bandgap harmonics. Instead our simulations indicate a transition from a prevailing intraband mechanism at \SI{0.5}{\tera\watt\per\centi\meter\squared} to a dominating interband mechanism at a peak intensity of \SI{10}{\tera\watt\per\centi\meter\squared}, while both contributions lead to comparable harmonic yields in between. \\
\\
Figures~\ref{fig:fig_5}(b)-(e) show the orientation-dependence of H3-H9 for four different SWIR peak intensities (\SIlist{0.5;1;5;10}{\tera\watt\per\centi\meter\squared}). The angular distribution of H3 changes from a four-fold symmetry with maxima at $\theta=\,$\SIlist{0;90;180;270}{\degree} (diatomic nearest-neighbour directions) at $I_0 =\,$\SI{0.5}{\tera\watt\per\centi\meter\squared} to an eight-fold symmetry at $I_0 =\,$\SI{1}{\tera\watt\per\centi\meter\squared}. At even higher SWIR peak intensities a four-fold symmetry with maxima at $\theta=\,$\SIlist{45;135;225;315}{\degree} (monoatomic nearest-neighbour directions) develops. Comparing these observations to Fig.~\ref{fig:fig_5}(a) indicates that the $\theta-$dependence associated with a dominating intraband mechanism (at $I_0 =\,$\SI{0.5}{\tera\watt\per\centi\meter\squared}, four-fold symmetry with maxima along diatomic nearest-neighbour directions) differs from the angular dependence due to the interband polarization (at $I_0 =\,$\SI{10}{\tera\watt\per\centi\meter\squared}, four-fold symmetry with maxima along monoatomic nearest-neighbour directions). Thus, the transition from a four-fold to an eight-fold symmetry corresponds to a transition from a purely intraband case to an intermediate situation where the combination of the interband polarization and the intraband current generates an eight-fold symmetry of the total H3 signal. The experimentally observed transition from a four-fold to an eight-fold symmetry of H3 (compare Fig.~\ref{fig:fig_2}) could hence be interpreted as a result of such a combined emission.\\
The $\theta$-dependence of H5 exhibits an eight-fold symmetry over the full range of SWIR intensities with one exception at $I_0=\,$\SI{5}{\tera\watt\per\centi\meter\squared} where the maxima along the diatomic nearest-neighbour directions break up into two peaks leading to a twelve-fold symmetry that was experimentally not observed. However, as the excitation intensity further increases up to \SI{10}{\tera\watt\per\centi\meter\squared}, an eight-fold symmetry in good agreement with our measurements is observed again. In the intensity range where we could experimentally access the seventh harmonic, the numerical simulations based on the SBEs reproduce the constant eight-fold symmetry of H7 while H9 tends to display a four-fold symmetric angular dependence whose preferred orientation for efficient HHG varies between the diatomic and the monoatomic nearest-neighbour directions [see Fig.~\ref{fig:fig_5}(e)]. \\
Generally, several microscopic physical mechanisms contribute to the two observables [$P(\omega)$ and $J(\omega)$] in the SBE-calculations. While interband recollisions are expected to dominate the above-bandgap response of the interband polarization $P(\omega)$, high-harmonic generation due to the intraband current is usually associated with quasi-free carriers being accelerated to non-parabolic regions of the energy bands. However, other HHG mechanisms that cannot easily be isolated are naturally included in these two observables. In particular, ionization-related HHG mechanisms are expected to severely contaminate the low-order harmonics \cite{Brunel_1990,Vampa_2017}. We expect the Brunel mechanism that is traditionally associated with harmonic generation due to photoionization in gases, clusters and solids to contribute to the intraband current as it relies upon carrier motion within the excited state. The injection current, in contrast, is directly connected to the spatial displacement of an electron during the interband excitation itself and is thus expected to contribute to the interband polarization.
To investigate a potential contribution of these ionization-related HHG mechanisms that cannot directly be extracted from the SBE calculations we performed further numerical simulations of the angular dependence of the photoionization rate. We have conceived a simple model for the photoionization rate that can support physical insights and provide realistic qualitative estimates of the angular dependence of solid-state HHG. Conventional computations of the $\theta$-dependence of the photoionization rate only consider carrier dynamics within distinct energy bands, i.e. transitions between Bloch-states that are based on single-atom wavefunctions localized at the same atom. Here we work in the Wannier-basis \cite{Osika_2017, Parks_2020, Brown_2022} which allows to explicitly resolve transitions from one atomic site to another. Since such nearest-neighbour transitions are not included in the dipole momenta used for the SBE calculations we treat them separately and qualitatively [see Fig.~\ref{fig:fig_6}(a)]. The photoionization rate (i.e. the rate of transitions from the valence to the conduction band) is approximated by the multiphoton formalism \cite{Ambrosek_2003} as
\begin{equation} \label{eq:mpi_rate}
\Gamma \approx \sum_s |\vec{E} \cdot \vec{s}|^{2N}d_{s}^{2N} 
\end{equation}
where the sum is taken over the neighbouring atomic sites, $\vec{E}\cdot\vec{s}$ is the projection of the electric field on the direction towards the atomic site, and $d_s$ is the transition dipole moment between a state in the valence zone and a state in the conduction zone at atomic site $s$. The electric field modifies the energy difference beetween these states and correspondingly the number of photons $N$ that is needed for a multiphoton transition. \\
For MgO we calculated the transition dipole moment based on the known energy structure of the isolated atoms, the bandgap value and the interatomic distances. For the MgO crystal the valence-zone electrons are localized at oxygen atoms, while the localized electrons in the conduction zone are predominantly positioned at oxygen atoms and partially at magnesium atoms \cite{DeBoer_1998}. For the calculations of the angle-dependent photoionization rate we have taken nearest-neighbour O-O transitions (corresponding to $\theta=\,$\SIlist{45;135;225;315}{\degree}), second-nearest-neighbour O-O transitions (corresponding to $\theta=\,$\SIlist{0;90;180;270}{\degree}) and nearest-neighbour O-Mg transitions (corresponding to $\theta=\,$\SIlist{0;90;180;270}{\degree}) into account [see Fig.~\ref{fig:fig_6}(b)-(d)]. We only analyze the $\theta$-dependence of the photoionization rate itself since all ionization-related harmonic orders are proportional to this rate and will inherit the same angular distribution. Numerical results for the $\theta$-dependence of the ionization rate are presented in Fig.~\ref{fig:fig_6}(b). At a low SWIR intensity of \SI{2}{\tera\watt\per\centi\meter\squared} a four-fold symmetry with maxima along the monoatomic nearest-neighbour directions is observed. This corresponds to the same directions for which the interband polarization dominates H3 [see brown line for $I_0=\,$\SI{10}{\tera\watt\per\centi\meter\squared} in Fig.~\ref{fig:fig_5}(b)]. As the excitation intensity increases, a transition from a four-fold to an eight-fold symmetry is found in the $\theta$-dependence of the photoionization rate [see orange line for a peak intensity of \SI{13}{\tera\watt\per\centi\meter\squared} in Fig.~\ref{fig:fig_6}(b)]. The additional maxima appearing at $\theta=\,$\SIlist{0;90;180;270}{\degree} can be associated with ionization along the diatomic nearest-neighbour directions (excitation from an O ion to a Mg ion or vice versa). Since this is a direct signature of Wannier-transitions from states localized at one ion to states localized at another ion, the SBE results do not include these maxima in the interband polarization [see Fig.~\ref{fig:fig_6}(a)]. The combination of both numerical approaches indicates that the interband polarization becomes dominant for H3 at sufficiently high intensities ($\geq\,$\SI{10}{\tera\watt\per\centi\meter\squared}), while the experimentally observed eight-fold symmetry is linked to Wannier-transitions between adjacent atoms. H3 (photon energy: \SI{2.48}{\electronvolt}) is situated well below the bandgap of bulk MgO ($\sim$\SI{7.8}{\electronvolt}), hence interband recollision can not be responsible for the detected emission. As the injection current is the only mechanism that generates below-bandgap harmonics in a non-perturbative manner and contributes to the interbant polarization, we attribute the emission of H3 at SWIR intensities $\geq\,$\SI{10}{\tera\watt\per\centi\meter\squared} to the injection current.
\section{Conclusion}
In summary, we have presented experimental and numerical results of the angular dependence of HHG in periodic crystals. To correlate the nonlinear frequency conversion process to laser-induced ionization we performed simultaneous transmission measurements that provide information on the ionization yield at different crystallographic orientations. Our investigations were focused on bulk MgO crystals with a cubic crystal structure and were complemented by experimental investigations in Al$_2$O$_3$ and LiF. In all materials a distinct correlation between HHG and nonlinear ionization was observed which manifested itself in a well-defined orientation dependence of both signals. In detail, we observed maxima of the harmonic emission coinciding with maxima of the nonlinear absorption. We interpreted this as an enhanced HHG conversion efficiency at angles where laser-induced ionization is also maximized. A symmetry analysis of the angular dependence of the experimentally observed signals connected the angles of maximum HHG emission and SWIR absorption with monoatomic and diatomic nearest-neighbour directions in the crystal lattice. \\
To further substantiate our interpretation and investigate the dominant mechanisms responsible for HHG we performed two sets of numerical simulations using MgO as a model system. First, we numerically solved the SBEs using one valence and one conduction band. Our results reproduced the experimentally observed transition from a four-fold to an eight-fold symmetry of H3 at intermediate intensity while at high intensity the interband polarization was found to dominate, resulting in a four-fold symmetry with maxima along the monoatomic nearest-neighbour directions. For H5 and H7 an eight-fold symmetry was predicted for the experimentally used SWIR intensities in agreement with our experimental findings. Within the SBE calculations the transition in the angular distribution was caused by a change from a dominating intraband current at low intensities to a prevailing interband polarization at higher intensities. The observed shift in dominant amplitude from  $A_{\rm{dia}}$  to $A_{\rm{mono}}$ in Al$_2$O$_3$ may also be attributed to the same phenomenon.  \\
Second, we numerically investigated the possible influence of ionization-related HHG mechanisms on the experimentally detected harmonic spectra. Our results unveiled a transition from a four-fold to an eight-fold symmetry of the photoionization rate for increasing intensity where the strongest signal is predicted along the monoatomic nearest neighbour directions equivalent to the dominant interband polarization contribution predicted by the SBEs. As Wannier-jumping (i.e. the transition from one atom to a neighbouring atom) is not included in the dipole momenta used in the SBE calculations the additionally emerging peaks leading to the eight-fold symmetry cannot be seen in the SBE results. Since the injection current is directly associated with interband excitations and thus contributes to $P(\omega)$ and since the recollision mechanism can be excluded due to the below-bandgap photon energy, we attribute - like in our previous work \cite{Juergens_2020, Juergens_2022} - the emission of H3 in the strong-field regime to the injection mechanism, the emission of harmonics due to the spatial displacement of electrons during the excitation process. \\

\section*{Funding.}
Funding by the German Research Foundation - SFB1477 "Light-Matter Interaction at Interfaces," Project No. 441234705, is gratefully acknowledged. 
The work of S.D.C.R.A. and P.M.K. has been carried out at the Advanced Research Center for Nanolithography (ARCNL), a public-private partnership of the University of Amsterdam (UvA), the Vrije Universiteit Amsterdam (VU), the Dutch Research Council (NWO), and the semiconductor equipment manufacturer ASML, and was partly financed by Toeslag voor Topconsortia voor Kennis en Innovatie (TKI) from the Dutch Ministry of Economic Affairs and Climate Policy. P.M.K. acknowledges support from ERC Starting Grant ANACONDA (grant no. 101041819).
\section*{Disclosures.}
The authors declare no conflicts of interest.
\section*{Acknowledgements}
It is our pleasant duty to thank M. Ivanov and T. Fennel for discussion of our experimental results and the numerical methods.

\section*{Data availablity.}
Data underlying the results presented in this paper are not publicly available at this time, but may be obtained from the authors upon reasonable request.

\bibliography{HHG_vs_SFI}

\begin{thebibliography}{44}%
\makeatletter
\providecommand \@ifxundefined [1]{%
 \@ifx{#1\undefined}
}%
\providecommand \@ifnum [1]{%
 \ifnum #1\expandafter \@firstoftwo
 \else \expandafter \@secondoftwo
 \fi
}%
\providecommand \@ifx [1]{%
 \ifx #1\expandafter \@firstoftwo
 \else \expandafter \@secondoftwo
 \fi
}%
\providecommand \natexlab [1]{#1}%
\providecommand \enquote  [1]{``#1''}%
\providecommand \bibnamefont  [1]{#1}%
\providecommand \bibfnamefont [1]{#1}%
\providecommand \citenamefont [1]{#1}%
\providecommand \href@noop [0]{\@secondoftwo}%
\providecommand \href [0]{\begingroup \@sanitize@url \@href}%
\providecommand \@href[1]{\@@startlink{#1}\@@href}%
\providecommand \@@href[1]{\endgroup#1\@@endlink}%
\providecommand \@sanitize@url [0]{\catcode `\\12\catcode `\$12\catcode
  `\&12\catcode `\#12\catcode `\^12\catcode `\_12\catcode `\%12\relax}%
\providecommand \@@startlink[1]{}%
\providecommand \@@endlink[0]{}%
\providecommand \url  [0]{\begingroup\@sanitize@url \@url }%
\providecommand \@url [1]{\endgroup\@href {#1}{\urlprefix }}%
\providecommand \urlprefix  [0]{URL }%
\providecommand \Eprint [0]{\href }%
\providecommand \doibase [0]{https://doi.org/}%
\providecommand \selectlanguage [0]{\@gobble}%
\providecommand \bibinfo  [0]{\@secondoftwo}%
\providecommand \bibfield  [0]{\@secondoftwo}%
\providecommand \translation [1]{[#1]}%
\providecommand \BibitemOpen [0]{}%
\providecommand \bibitemStop [0]{}%
\providecommand \bibitemNoStop [0]{.\EOS\space}%
\providecommand \EOS [0]{\spacefactor3000\relax}%
\providecommand \BibitemShut  [1]{\csname bibitem#1\endcsname}%
\let\auto@bib@innerbib\@empty
\bibitem [{\citenamefont {Corkum}(1993)}]{Corkum_1993}%
  \BibitemOpen
  \bibfield  {author} {\bibinfo {author} {\bibfnamefont {P.~B.}\ \bibnamefont
  {Corkum}},\ }\bibfield  {title} {\bibinfo {title} {Plasma perspective on
  strong field multiphoton ionization},\ }\href@noop {} {\bibfield  {journal}
  {\bibinfo  {journal} {Physical Review Letters}\ }\textbf {\bibinfo {volume}
  {71}},\ \bibinfo {pages} {1994} (\bibinfo {year} {1993})}\BibitemShut
  {NoStop}%
\bibitem [{\citenamefont {Lewenstein}\ \emph {et~al.}(1994)\citenamefont
  {Lewenstein}, \citenamefont {Balcou}, \citenamefont {Ivanov}, \citenamefont
  {L’Huillier},\ and\ \citenamefont {Corkum}}]{Lewenstein_1994}%
  \BibitemOpen
  \bibfield  {author} {\bibinfo {author} {\bibfnamefont {M.}~\bibnamefont
  {Lewenstein}}, \bibinfo {author} {\bibfnamefont {P.}~\bibnamefont {Balcou}},
  \bibinfo {author} {\bibfnamefont {M.~Y.}\ \bibnamefont {Ivanov}}, \bibinfo
  {author} {\bibfnamefont {A.}~\bibnamefont {L’Huillier}},\ and\ \bibinfo
  {author} {\bibfnamefont {P.~B.}\ \bibnamefont {Corkum}},\ }\bibfield  {title}
  {\bibinfo {title} {Theory of high-harmonic generation by low-frequency laser
  fields},\ }\href@noop {} {\bibfield  {journal} {\bibinfo  {journal} {Physical
  Review A}\ }\textbf {\bibinfo {volume} {49}},\ \bibinfo {pages} {2117}
  (\bibinfo {year} {1994})}\BibitemShut {NoStop}%
\bibitem [{\citenamefont {Krausz}\ and\ \citenamefont
  {Ivanov}(2009)}]{Krausz_2009}%
  \BibitemOpen
  \bibfield  {author} {\bibinfo {author} {\bibfnamefont {F.}~\bibnamefont
  {Krausz}}\ and\ \bibinfo {author} {\bibfnamefont {M.}~\bibnamefont
  {Ivanov}},\ }\bibfield  {title} {\bibinfo {title} {Attosecond physics},\
  }\href@noop {} {\bibfield  {journal} {\bibinfo  {journal} {Reviews of Modern
  Physics}\ }\textbf {\bibinfo {volume} {81}},\ \bibinfo {pages} {163}
  (\bibinfo {year} {2009})}\BibitemShut {NoStop}%
\bibitem [{\citenamefont {Sansone}\ \emph {et~al.}(2010)\citenamefont
  {Sansone}, \citenamefont {Kelkensberg}, \citenamefont {P{\'e}rez-Torres},
  \citenamefont {Morales}, \citenamefont {Kling}, \citenamefont {Siu},
  \citenamefont {Ghafur}, \citenamefont {Johnsson}, \citenamefont {Swoboda},
  \citenamefont {Benedetti} \emph {et~al.}}]{Sansone_2010}%
  \BibitemOpen
  \bibfield  {author} {\bibinfo {author} {\bibfnamefont {G.}~\bibnamefont
  {Sansone}}, \bibinfo {author} {\bibfnamefont {F.}~\bibnamefont
  {Kelkensberg}}, \bibinfo {author} {\bibfnamefont {J.}~\bibnamefont
  {P{\'e}rez-Torres}}, \bibinfo {author} {\bibfnamefont {F.}~\bibnamefont
  {Morales}}, \bibinfo {author} {\bibfnamefont {M.~F.}\ \bibnamefont {Kling}},
  \bibinfo {author} {\bibfnamefont {W.}~\bibnamefont {Siu}}, \bibinfo {author}
  {\bibfnamefont {O.}~\bibnamefont {Ghafur}}, \bibinfo {author} {\bibfnamefont
  {P.}~\bibnamefont {Johnsson}}, \bibinfo {author} {\bibfnamefont
  {M.}~\bibnamefont {Swoboda}}, \bibinfo {author} {\bibfnamefont
  {E.}~\bibnamefont {Benedetti}}, \emph {et~al.},\ }\bibfield  {title}
  {\bibinfo {title} {Electron localization following attosecond molecular
  photoionization},\ }\href@noop {} {\bibfield  {journal} {\bibinfo  {journal}
  {Nature}\ }\textbf {\bibinfo {volume} {465}},\ \bibinfo {pages} {763}
  (\bibinfo {year} {2010})}\BibitemShut {NoStop}%
\bibitem [{\citenamefont {Ghimire}\ \emph {et~al.}(2011)\citenamefont
  {Ghimire}, \citenamefont {DiChiara}, \citenamefont {Sistrunk}, \citenamefont
  {Agostini}, \citenamefont {DiMauro},\ and\ \citenamefont
  {Reis}}]{Ghimire_2011}%
  \BibitemOpen
  \bibfield  {author} {\bibinfo {author} {\bibfnamefont {S.}~\bibnamefont
  {Ghimire}}, \bibinfo {author} {\bibfnamefont {A.~D.}\ \bibnamefont
  {DiChiara}}, \bibinfo {author} {\bibfnamefont {E.}~\bibnamefont {Sistrunk}},
  \bibinfo {author} {\bibfnamefont {P.}~\bibnamefont {Agostini}}, \bibinfo
  {author} {\bibfnamefont {L.~F.}\ \bibnamefont {DiMauro}},\ and\ \bibinfo
  {author} {\bibfnamefont {D.~A.}\ \bibnamefont {Reis}},\ }\bibfield  {title}
  {\bibinfo {title} {Observation of high-order harmonic generation in a bulk
  crystal},\ }\href@noop {} {\bibfield  {journal} {\bibinfo  {journal} {Nature
  Physics}\ }\textbf {\bibinfo {volume} {7}},\ \bibinfo {pages} {138} (\bibinfo
  {year} {2011})}\BibitemShut {NoStop}%
\bibitem [{\citenamefont {Schubert}\ \emph {et~al.}(2014)\citenamefont
  {Schubert}, \citenamefont {Hohenleutner}, \citenamefont {Langer},
  \citenamefont {Urbanek}, \citenamefont {Lange}, \citenamefont {Huttner},
  \citenamefont {Golde}, \citenamefont {Meier}, \citenamefont {Kira},
  \citenamefont {Koch} \emph {et~al.}}]{Schubert_2014}%
  \BibitemOpen
  \bibfield  {author} {\bibinfo {author} {\bibfnamefont {O.}~\bibnamefont
  {Schubert}}, \bibinfo {author} {\bibfnamefont {M.}~\bibnamefont
  {Hohenleutner}}, \bibinfo {author} {\bibfnamefont {F.}~\bibnamefont
  {Langer}}, \bibinfo {author} {\bibfnamefont {B.}~\bibnamefont {Urbanek}},
  \bibinfo {author} {\bibfnamefont {C.}~\bibnamefont {Lange}}, \bibinfo
  {author} {\bibfnamefont {U.}~\bibnamefont {Huttner}}, \bibinfo {author}
  {\bibfnamefont {D.}~\bibnamefont {Golde}}, \bibinfo {author} {\bibfnamefont
  {T.}~\bibnamefont {Meier}}, \bibinfo {author} {\bibfnamefont
  {M.}~\bibnamefont {Kira}}, \bibinfo {author} {\bibfnamefont {S.~W.}\
  \bibnamefont {Koch}}, \emph {et~al.},\ }\bibfield  {title} {\bibinfo {title}
  {{Sub-cycle control of terahertz high-harmonic generation by dynamical Bloch
  oscillations}},\ }\href@noop {} {\bibfield  {journal} {\bibinfo  {journal}
  {Nature Photonics}\ }\textbf {\bibinfo {volume} {8}},\ \bibinfo {pages} {119}
  (\bibinfo {year} {2014})}\BibitemShut {NoStop}%
\bibitem [{\citenamefont {Garg}\ \emph {et~al.}(2018)\citenamefont {Garg},
  \citenamefont {Kim},\ and\ \citenamefont {Goulielmakis}}]{Garg_2018}%
  \BibitemOpen
  \bibfield  {author} {\bibinfo {author} {\bibfnamefont {M.}~\bibnamefont
  {Garg}}, \bibinfo {author} {\bibfnamefont {H.-Y.}\ \bibnamefont {Kim}},\ and\
  \bibinfo {author} {\bibfnamefont {E.}~\bibnamefont {Goulielmakis}},\
  }\bibfield  {title} {\bibinfo {title} {Ultimate waveform reproducibility of
  extreme-ultraviolet pulses by high-harmonic generation in quartz},\
  }\href@noop {} {\bibfield  {journal} {\bibinfo  {journal} {Nature Photonics}\
  }\textbf {\bibinfo {volume} {12}},\ \bibinfo {pages} {291} (\bibinfo {year}
  {2018})}\BibitemShut {NoStop}%
\bibitem [{\citenamefont {Garg}\ \emph {et~al.}(2016)\citenamefont {Garg},
  \citenamefont {Zhan}, \citenamefont {Luu}, \citenamefont {Lakhotia},
  \citenamefont {Klostermann}, \citenamefont {Guggenmos},\ and\ \citenamefont
  {Goulielmakis}}]{Garg_2016}%
  \BibitemOpen
  \bibfield  {author} {\bibinfo {author} {\bibfnamefont {M.}~\bibnamefont
  {Garg}}, \bibinfo {author} {\bibfnamefont {M.}~\bibnamefont {Zhan}}, \bibinfo
  {author} {\bibfnamefont {T.~T.}\ \bibnamefont {Luu}}, \bibinfo {author}
  {\bibfnamefont {H.}~\bibnamefont {Lakhotia}}, \bibinfo {author}
  {\bibfnamefont {T.}~\bibnamefont {Klostermann}}, \bibinfo {author}
  {\bibfnamefont {A.}~\bibnamefont {Guggenmos}},\ and\ \bibinfo {author}
  {\bibfnamefont {E.}~\bibnamefont {Goulielmakis}},\ }\bibfield  {title}
  {\bibinfo {title} {Multi-petahertz electronic metrology},\ }\href@noop {}
  {\bibfield  {journal} {\bibinfo  {journal} {Nature}\ }\textbf {\bibinfo
  {volume} {538}},\ \bibinfo {pages} {359} (\bibinfo {year}
  {2016})}\BibitemShut {NoStop}%
\bibitem [{\citenamefont {Ossiander}\ \emph {et~al.}(2022)\citenamefont
  {Ossiander}, \citenamefont {Golyari}, \citenamefont {Scharl}, \citenamefont
  {Lehnert}, \citenamefont {Siegrist}, \citenamefont {B{\"u}rger},
  \citenamefont {Zimin}, \citenamefont {Gessner}, \citenamefont {Weidman},
  \citenamefont {Floss} \emph {et~al.}}]{Ossiander_2022}%
  \BibitemOpen
  \bibfield  {author} {\bibinfo {author} {\bibfnamefont {M.}~\bibnamefont
  {Ossiander}}, \bibinfo {author} {\bibfnamefont {K.}~\bibnamefont {Golyari}},
  \bibinfo {author} {\bibfnamefont {K.}~\bibnamefont {Scharl}}, \bibinfo
  {author} {\bibfnamefont {L.}~\bibnamefont {Lehnert}}, \bibinfo {author}
  {\bibfnamefont {F.}~\bibnamefont {Siegrist}}, \bibinfo {author}
  {\bibfnamefont {J.}~\bibnamefont {B{\"u}rger}}, \bibinfo {author}
  {\bibfnamefont {D.}~\bibnamefont {Zimin}}, \bibinfo {author} {\bibfnamefont
  {J.}~\bibnamefont {Gessner}}, \bibinfo {author} {\bibfnamefont
  {M.}~\bibnamefont {Weidman}}, \bibinfo {author} {\bibfnamefont
  {I.}~\bibnamefont {Floss}}, \emph {et~al.},\ }\bibfield  {title} {\bibinfo
  {title} {The speed limit of optoelectronics},\ }\href@noop {} {\bibfield
  {journal} {\bibinfo  {journal} {Nature Communications}\ }\textbf {\bibinfo
  {volume} {13}},\ \bibinfo {pages} {1} (\bibinfo {year} {2022})}\BibitemShut
  {NoStop}%
\bibitem [{\citenamefont {Stolen}\ and\ \citenamefont
  {Ashkin}(1973)}]{Stolen_1973}%
  \BibitemOpen
  \bibfield  {author} {\bibinfo {author} {\bibfnamefont {R.}~\bibnamefont
  {Stolen}}\ and\ \bibinfo {author} {\bibfnamefont {A.}~\bibnamefont
  {Ashkin}},\ }\bibfield  {title} {\bibinfo {title} {{Optical Kerr effect in
  glass waveguide}},\ }\href@noop {} {\bibfield  {journal} {\bibinfo  {journal}
  {Applied Physics Letters}\ }\textbf {\bibinfo {volume} {22}},\ \bibinfo
  {pages} {294} (\bibinfo {year} {1973})}\BibitemShut {NoStop}%
\bibitem [{\citenamefont {Franken}\ \emph {et~al.}(1961)\citenamefont
  {Franken}, \citenamefont {Hill}, \citenamefont {Peters},\ and\ \citenamefont
  {Weinreich}}]{Franken_1961}%
  \BibitemOpen
  \bibfield  {author} {\bibinfo {author} {\bibfnamefont {P.~A.}\ \bibnamefont
  {Franken}}, \bibinfo {author} {\bibfnamefont {A.~E.}\ \bibnamefont {Hill}},
  \bibinfo {author} {\bibfnamefont {C.~W.}\ \bibnamefont {Peters}},\ and\
  \bibinfo {author} {\bibfnamefont {G.}~\bibnamefont {Weinreich}},\ }\bibfield
  {title} {\bibinfo {title} {Generation of optical harmonics},\ }\href@noop {}
  {\bibfield  {journal} {\bibinfo  {journal} {Physical Review Letters}\
  }\textbf {\bibinfo {volume} {7}},\ \bibinfo {pages} {118} (\bibinfo {year}
  {1961})}\BibitemShut {NoStop}%
\bibitem [{\citenamefont {Elu}\ \emph {et~al.}(2017)\citenamefont {Elu},
  \citenamefont {Baudisch}, \citenamefont {Pires}, \citenamefont {Tani},
  \citenamefont {Frosz}, \citenamefont {K{\"o}ttig}, \citenamefont {Ermolov},
  \citenamefont {Russell},\ and\ \citenamefont {Biegert}}]{Elu_2017}%
  \BibitemOpen
  \bibfield  {author} {\bibinfo {author} {\bibfnamefont {U.}~\bibnamefont
  {Elu}}, \bibinfo {author} {\bibfnamefont {M.}~\bibnamefont {Baudisch}},
  \bibinfo {author} {\bibfnamefont {H.}~\bibnamefont {Pires}}, \bibinfo
  {author} {\bibfnamefont {F.}~\bibnamefont {Tani}}, \bibinfo {author}
  {\bibfnamefont {M.~H.}\ \bibnamefont {Frosz}}, \bibinfo {author}
  {\bibfnamefont {F.}~\bibnamefont {K{\"o}ttig}}, \bibinfo {author}
  {\bibfnamefont {A.}~\bibnamefont {Ermolov}}, \bibinfo {author} {\bibfnamefont
  {P.~S.~J.}\ \bibnamefont {Russell}},\ and\ \bibinfo {author} {\bibfnamefont
  {J.}~\bibnamefont {Biegert}},\ }\bibfield  {title} {\bibinfo {title} {{High
  average power and single-cycle pulses from a mid-IR optical parametric
  chirped pulse amplifier}},\ }\href@noop {} {\bibfield  {journal} {\bibinfo
  {journal} {Optica}\ }\textbf {\bibinfo {volume} {4}},\ \bibinfo {pages}
  {1024} (\bibinfo {year} {2017})}\BibitemShut {NoStop}%
\bibitem [{\citenamefont {Luu}\ \emph {et~al.}(2015)\citenamefont {Luu},
  \citenamefont {Garg}, \citenamefont {Kruchinin}, \citenamefont {Moulet},
  \citenamefont {Hassan},\ and\ \citenamefont {Goulielmakis}}]{Luu_2015}%
  \BibitemOpen
  \bibfield  {author} {\bibinfo {author} {\bibfnamefont {T.~T.}\ \bibnamefont
  {Luu}}, \bibinfo {author} {\bibfnamefont {M.}~\bibnamefont {Garg}}, \bibinfo
  {author} {\bibfnamefont {S.~Y.}\ \bibnamefont {Kruchinin}}, \bibinfo {author}
  {\bibfnamefont {A.}~\bibnamefont {Moulet}}, \bibinfo {author} {\bibfnamefont
  {M.~T.}\ \bibnamefont {Hassan}},\ and\ \bibinfo {author} {\bibfnamefont
  {E.}~\bibnamefont {Goulielmakis}},\ }\bibfield  {title} {\bibinfo {title}
  {Extreme ultraviolet high-harmonic spectroscopy of solids},\ }\href@noop {}
  {\bibfield  {journal} {\bibinfo  {journal} {Nature}\ }\textbf {\bibinfo
  {volume} {521}},\ \bibinfo {pages} {498} (\bibinfo {year}
  {2015})}\BibitemShut {NoStop}%
\bibitem [{\citenamefont {Krause}\ \emph {et~al.}(1992)\citenamefont {Krause},
  \citenamefont {Schafer},\ and\ \citenamefont {Kulander}}]{Krause_1992}%
  \BibitemOpen
  \bibfield  {author} {\bibinfo {author} {\bibfnamefont {J.~L.}\ \bibnamefont
  {Krause}}, \bibinfo {author} {\bibfnamefont {K.~J.}\ \bibnamefont
  {Schafer}},\ and\ \bibinfo {author} {\bibfnamefont {K.~C.}\ \bibnamefont
  {Kulander}},\ }\bibfield  {title} {\bibinfo {title} {High-order harmonic
  generation from atoms and ions in the high intensity regime},\ }\href@noop {}
  {\bibfield  {journal} {\bibinfo  {journal} {Physical Review Letters}\
  }\textbf {\bibinfo {volume} {68}},\ \bibinfo {pages} {3535} (\bibinfo {year}
  {1992})}\BibitemShut {NoStop}%
\bibitem [{\citenamefont {Goulielmakis}\ and\ \citenamefont
  {Brabec}(2022)}]{Goulielmakis_2022}%
  \BibitemOpen
  \bibfield  {author} {\bibinfo {author} {\bibfnamefont {E.}~\bibnamefont
  {Goulielmakis}}\ and\ \bibinfo {author} {\bibfnamefont {T.}~\bibnamefont
  {Brabec}},\ }\bibfield  {title} {\bibinfo {title} {High harmonic generation
  in condensed matter},\ }\href@noop {} {\bibfield  {journal} {\bibinfo
  {journal} {Nature Photonics}\ }\textbf {\bibinfo {volume} {16}},\ \bibinfo
  {pages} {411} (\bibinfo {year} {2022})}\BibitemShut {NoStop}%
\bibitem [{\citenamefont {Brunel}(1990)}]{Brunel_1990}%
  \BibitemOpen
  \bibfield  {author} {\bibinfo {author} {\bibfnamefont {F.}~\bibnamefont
  {Brunel}},\ }\bibfield  {title} {\bibinfo {title} {Harmonic generation due to
  plasma effects in a gas undergoing multiphoton ionization in the
  high-intensity limit},\ }\href@noop {} {\bibfield  {journal} {\bibinfo
  {journal} {JOSA B}\ }\textbf {\bibinfo {volume} {7}},\ \bibinfo {pages} {521}
  (\bibinfo {year} {1990})}\BibitemShut {NoStop}%
\bibitem [{\citenamefont {Bauer}\ \emph {et~al.}(1998)\citenamefont {Bauer},
  \citenamefont {Salomaa},\ and\ \citenamefont {Mulser}}]{Bauer_1998}%
  \BibitemOpen
  \bibfield  {author} {\bibinfo {author} {\bibfnamefont {D.}~\bibnamefont
  {Bauer}}, \bibinfo {author} {\bibfnamefont {R.}~\bibnamefont {Salomaa}},\
  and\ \bibinfo {author} {\bibfnamefont {P.}~\bibnamefont {Mulser}},\
  }\bibfield  {title} {\bibinfo {title} {Generation of ultrashort light pulses
  by a rapidly ionizing thin foil},\ }\href@noop {} {\bibfield  {journal}
  {\bibinfo  {journal} {Physical Review E}\ }\textbf {\bibinfo {volume} {58}},\
  \bibinfo {pages} {2436} (\bibinfo {year} {1998})}\BibitemShut {NoStop}%
\bibitem [{\citenamefont {{Geissler, Michael and Tempea, Gabriel and Scrinzi,
  Armin and Schn{\"u}rer, Matthias and Krausz, Ferenc and Brabec,
  Thomas}}(1999)}]{Geissler_1999}%
  \BibitemOpen
  \bibfield  {author} {\bibinfo {author} {\bibnamefont {{Geissler, Michael and
  Tempea, Gabriel and Scrinzi, Armin and Schn{\"u}rer, Matthias and Krausz,
  Ferenc and Brabec, Thomas}}},\ }\bibfield  {title} {\bibinfo {title} {Light
  propagation in field-ionizing media: extreme nonlinear optics},\ }\href@noop
  {} {\bibfield  {journal} {\bibinfo  {journal} {{Physical Review Letters}}\
  }\textbf {\bibinfo {volume} {83}},\ \bibinfo {pages} {2930} (\bibinfo {year}
  {1999})}\BibitemShut {NoStop}%
\bibitem [{\citenamefont {J{\"u}rgens}\ \emph {et~al.}(2022)\citenamefont
  {J{\"u}rgens}, \citenamefont {Liewehr}, \citenamefont {Kruse}, \citenamefont
  {Peltz}, \citenamefont {Witting}, \citenamefont {Husakou}, \citenamefont
  {Rouzee}, \citenamefont {Ivanov}, \citenamefont {Fennel}, \citenamefont
  {Vrakking} \emph {et~al.}}]{Juergens_2022}%
  \BibitemOpen
  \bibfield  {author} {\bibinfo {author} {\bibfnamefont {P.}~\bibnamefont
  {J{\"u}rgens}}, \bibinfo {author} {\bibfnamefont {B.}~\bibnamefont
  {Liewehr}}, \bibinfo {author} {\bibfnamefont {B.}~\bibnamefont {Kruse}},
  \bibinfo {author} {\bibfnamefont {C.}~\bibnamefont {Peltz}}, \bibinfo
  {author} {\bibfnamefont {T.}~\bibnamefont {Witting}}, \bibinfo {author}
  {\bibfnamefont {A.}~\bibnamefont {Husakou}}, \bibinfo {author} {\bibfnamefont
  {A.}~\bibnamefont {Rouzee}}, \bibinfo {author} {\bibfnamefont
  {M.}~\bibnamefont {Ivanov}}, \bibinfo {author} {\bibfnamefont
  {T.}~\bibnamefont {Fennel}}, \bibinfo {author} {\bibfnamefont {M.~J.}\
  \bibnamefont {Vrakking}}, \emph {et~al.},\ }\bibfield  {title} {\bibinfo
  {title} {Characterization of laser-induced ionization dynamics in solid
  dielectrics},\ }\href@noop {} {\bibfield  {journal} {\bibinfo  {journal} {ACS
  Photonics}\ }\textbf {\bibinfo {volume} {9}},\ \bibinfo {pages} {233}
  (\bibinfo {year} {2022})}\BibitemShut {NoStop}%
\bibitem [{\citenamefont {Verhoef}\ \emph {et~al.}(2010)\citenamefont
  {Verhoef}, \citenamefont {Mitrofanov}, \citenamefont {Serebryannikov},
  \citenamefont {Kartashov}, \citenamefont {Zheltikov},\ and\ \citenamefont
  {Baltu{\v{s}}ka}}]{Verhoef_2010}%
  \BibitemOpen
  \bibfield  {author} {\bibinfo {author} {\bibfnamefont {A.~J.}\ \bibnamefont
  {Verhoef}}, \bibinfo {author} {\bibfnamefont {A.~V.}\ \bibnamefont
  {Mitrofanov}}, \bibinfo {author} {\bibfnamefont {E.~E.}\ \bibnamefont
  {Serebryannikov}}, \bibinfo {author} {\bibfnamefont {D.~V.}\ \bibnamefont
  {Kartashov}}, \bibinfo {author} {\bibfnamefont {A.~M.}\ \bibnamefont
  {Zheltikov}},\ and\ \bibinfo {author} {\bibfnamefont {A.}~\bibnamefont
  {Baltu{\v{s}}ka}},\ }\bibfield  {title} {\bibinfo {title} {Optical detection
  of tunneling ionization},\ }\href@noop {} {\bibfield  {journal} {\bibinfo
  {journal} {Physical Review Letters}\ }\textbf {\bibinfo {volume} {104}},\
  \bibinfo {pages} {163904} (\bibinfo {year} {2010})}\BibitemShut {NoStop}%
\bibitem [{\citenamefont {Babushkin}\ \emph {et~al.}(2022)\citenamefont
  {Babushkin}, \citenamefont {Gal{\'a}n}, \citenamefont {de~Andrade},
  \citenamefont {Husakou}, \citenamefont {Morales}, \citenamefont {Kretschmar},
  \citenamefont {Nagy}, \citenamefont {Vai{\v{c}}aitis}, \citenamefont {Shi},
  \citenamefont {Zuber} \emph {et~al.}}]{Babushkin_2022}%
  \BibitemOpen
  \bibfield  {author} {\bibinfo {author} {\bibfnamefont {I.}~\bibnamefont
  {Babushkin}}, \bibinfo {author} {\bibfnamefont {{\'A}.~J.}\ \bibnamefont
  {Gal{\'a}n}}, \bibinfo {author} {\bibfnamefont {J.~R.~C.}\ \bibnamefont
  {de~Andrade}}, \bibinfo {author} {\bibfnamefont {A.}~\bibnamefont {Husakou}},
  \bibinfo {author} {\bibfnamefont {F.}~\bibnamefont {Morales}}, \bibinfo
  {author} {\bibfnamefont {M.}~\bibnamefont {Kretschmar}}, \bibinfo {author}
  {\bibfnamefont {T.}~\bibnamefont {Nagy}}, \bibinfo {author} {\bibfnamefont
  {V.}~\bibnamefont {Vai{\v{c}}aitis}}, \bibinfo {author} {\bibfnamefont
  {L.}~\bibnamefont {Shi}}, \bibinfo {author} {\bibfnamefont {D.}~\bibnamefont
  {Zuber}}, \emph {et~al.},\ }\bibfield  {title} {\bibinfo {title} {All-optical
  attoclock for imaging tunnelling wavepackets},\ }\href@noop {} {\bibfield
  {journal} {\bibinfo  {journal} {Nature Physics}\ }\textbf {\bibinfo {volume}
  {18}},\ \bibinfo {pages} {417} (\bibinfo {year} {2022})}\BibitemShut
  {NoStop}%
\bibitem [{\citenamefont {Gao}\ \emph {et~al.}(2019)\citenamefont {Gao},
  \citenamefont {Shim},\ and\ \citenamefont {Downer}}]{Gao_2019}%
  \BibitemOpen
  \bibfield  {author} {\bibinfo {author} {\bibfnamefont {X.}~\bibnamefont
  {Gao}}, \bibinfo {author} {\bibfnamefont {B.}~\bibnamefont {Shim}},\ and\
  \bibinfo {author} {\bibfnamefont {M.~C.}\ \bibnamefont {Downer}},\ }\bibfield
   {title} {\bibinfo {title} {Brunel harmonics generated from ionizing clusters
  by few-cycle laser pulses},\ }\href@noop {} {\bibfield  {journal} {\bibinfo
  {journal} {Optics Letters}\ }\textbf {\bibinfo {volume} {44}},\ \bibinfo
  {pages} {779} (\bibinfo {year} {2019})}\BibitemShut {NoStop}%
\bibitem [{\citenamefont {Mitrofanov}\ \emph {et~al.}(2011)\citenamefont
  {Mitrofanov}, \citenamefont {Verhoef}, \citenamefont {Serebryannikov},
  \citenamefont {Lumeau}, \citenamefont {Glebov}, \citenamefont {Zheltikov},\
  and\ \citenamefont {Baltu{\v{s}}ka}}]{Mitrofanov_2011}%
  \BibitemOpen
  \bibfield  {author} {\bibinfo {author} {\bibfnamefont {A.~V.}\ \bibnamefont
  {Mitrofanov}}, \bibinfo {author} {\bibfnamefont {A.~J.}\ \bibnamefont
  {Verhoef}}, \bibinfo {author} {\bibfnamefont {E.~E.}\ \bibnamefont
  {Serebryannikov}}, \bibinfo {author} {\bibfnamefont {J.}~\bibnamefont
  {Lumeau}}, \bibinfo {author} {\bibfnamefont {L.}~\bibnamefont {Glebov}},
  \bibinfo {author} {\bibfnamefont {A.~M.}\ \bibnamefont {Zheltikov}},\ and\
  \bibinfo {author} {\bibfnamefont {A.}~\bibnamefont {Baltu{\v{s}}ka}},\
  }\bibfield  {title} {\bibinfo {title} {Optical detection of attosecond
  ionization induced by a few-cycle laser field in a transparent dielectric
  material},\ }\href@noop {} {\bibfield  {journal} {\bibinfo  {journal}
  {Physical Review Letters}\ }\textbf {\bibinfo {volume} {106}},\ \bibinfo
  {pages} {147401} (\bibinfo {year} {2011})}\BibitemShut {NoStop}%
\bibitem [{\citenamefont {Li}\ \emph {et~al.}(2022)\citenamefont {Li},
  \citenamefont {Saleh}, \citenamefont {Sharma}, \citenamefont {Sierka},
  \citenamefont {H{\"u}necke}, \citenamefont {Neuhaus}, \citenamefont
  {Hedewig}, \citenamefont {Bergues}, \citenamefont {Alharbi}, \citenamefont
  {Azzeer} \emph {et~al.}}]{Li_2022}%
  \BibitemOpen
  \bibfield  {author} {\bibinfo {author} {\bibfnamefont {W.}~\bibnamefont
  {Li}}, \bibinfo {author} {\bibfnamefont {A.}~\bibnamefont {Saleh}}, \bibinfo
  {author} {\bibfnamefont {M.}~\bibnamefont {Sharma}}, \bibinfo {author}
  {\bibfnamefont {M.}~\bibnamefont {Sierka}}, \bibinfo {author} {\bibfnamefont
  {C.}~\bibnamefont {H{\"u}necke}}, \bibinfo {author} {\bibfnamefont
  {M.}~\bibnamefont {Neuhaus}}, \bibinfo {author} {\bibfnamefont
  {L.}~\bibnamefont {Hedewig}}, \bibinfo {author} {\bibfnamefont
  {B.}~\bibnamefont {Bergues}}, \bibinfo {author} {\bibfnamefont
  {M.}~\bibnamefont {Alharbi}}, \bibinfo {author} {\bibfnamefont {A.~M.}\
  \bibnamefont {Azzeer}}, \emph {et~al.},\ }\bibfield  {title} {\bibinfo
  {title} {{Resonance effects in Brunel harmonic generation in thin film
  organic semiconductors}},\ }\href@noop {} {\bibfield  {journal} {\bibinfo
  {journal} {arXiv preprint arXiv:2211.07062}\ } (\bibinfo {year}
  {2022})}\BibitemShut {NoStop}%
\bibitem [{\citenamefont {J{\"u}rgens}\ \emph {et~al.}(2020)\citenamefont
  {J{\"u}rgens}, \citenamefont {Liewehr}, \citenamefont {Kruse}, \citenamefont
  {Peltz}, \citenamefont {Engel}, \citenamefont {Husakou}, \citenamefont
  {Witting}, \citenamefont {Ivanov}, \citenamefont {Vrakking}, \citenamefont
  {Fennel} \emph {et~al.}}]{Juergens_2020}%
  \BibitemOpen
  \bibfield  {author} {\bibinfo {author} {\bibfnamefont {P.}~\bibnamefont
  {J{\"u}rgens}}, \bibinfo {author} {\bibfnamefont {B.}~\bibnamefont
  {Liewehr}}, \bibinfo {author} {\bibfnamefont {B.}~\bibnamefont {Kruse}},
  \bibinfo {author} {\bibfnamefont {C.}~\bibnamefont {Peltz}}, \bibinfo
  {author} {\bibfnamefont {D.}~\bibnamefont {Engel}}, \bibinfo {author}
  {\bibfnamefont {A.}~\bibnamefont {Husakou}}, \bibinfo {author} {\bibfnamefont
  {T.}~\bibnamefont {Witting}}, \bibinfo {author} {\bibfnamefont
  {M.}~\bibnamefont {Ivanov}}, \bibinfo {author} {\bibfnamefont
  {M.}~\bibnamefont {Vrakking}}, \bibinfo {author} {\bibfnamefont
  {T.}~\bibnamefont {Fennel}}, \emph {et~al.},\ }\bibfield  {title} {\bibinfo
  {title} {Origin of strong-field-induced low-order harmonic generation in
  amorphous quartz},\ }\href@noop {} {\bibfield  {journal} {\bibinfo  {journal}
  {Nature Physics}\ }\textbf {\bibinfo {volume} {16}},\ \bibinfo {pages} {1035}
  (\bibinfo {year} {2020})}\BibitemShut {NoStop}%
\bibitem [{\citenamefont {Lanin}\ \emph {et~al.}(2019)\citenamefont {Lanin},
  \citenamefont {Stepanov}, \citenamefont {Mitrofanov}, \citenamefont
  {Sidorov-Biryukov}, \citenamefont {Fedotov},\ and\ \citenamefont
  {Zheltikov}}]{Lanin_2019}%
  \BibitemOpen
  \bibfield  {author} {\bibinfo {author} {\bibfnamefont {A.~A.}\ \bibnamefont
  {Lanin}}, \bibinfo {author} {\bibfnamefont {E.~A.}\ \bibnamefont {Stepanov}},
  \bibinfo {author} {\bibfnamefont {A.~V.}\ \bibnamefont {Mitrofanov}},
  \bibinfo {author} {\bibfnamefont {D.~A.}\ \bibnamefont {Sidorov-Biryukov}},
  \bibinfo {author} {\bibfnamefont {A.~B.}\ \bibnamefont {Fedotov}},\ and\
  \bibinfo {author} {\bibfnamefont {A.~M.}\ \bibnamefont {Zheltikov}},\
  }\bibfield  {title} {\bibinfo {title} {High-order harmonic analysis of
  anisotropic petahertz photocurrents in solids},\ }\href@noop {} {\bibfield
  {journal} {\bibinfo  {journal} {Optics Letters}\ }\textbf {\bibinfo {volume}
  {44}},\ \bibinfo {pages} {1888} (\bibinfo {year} {2019})}\BibitemShut
  {NoStop}%
\bibitem [{\citenamefont {You}\ \emph {et~al.}(2017)\citenamefont {You},
  \citenamefont {Reis},\ and\ \citenamefont {Ghimire}}]{You_2017}%
  \BibitemOpen
  \bibfield  {author} {\bibinfo {author} {\bibfnamefont {Y.~S.}\ \bibnamefont
  {You}}, \bibinfo {author} {\bibfnamefont {D.~A.}\ \bibnamefont {Reis}},\ and\
  \bibinfo {author} {\bibfnamefont {S.}~\bibnamefont {Ghimire}},\ }\bibfield
  {title} {\bibinfo {title} {Anisotropic high-harmonic generation in bulk
  crystals},\ }\href@noop {} {\bibfield  {journal} {\bibinfo  {journal} {Nature
  Physics}\ }\textbf {\bibinfo {volume} {13}},\ \bibinfo {pages} {345}
  (\bibinfo {year} {2017})}\BibitemShut {NoStop}%
\bibitem [{\citenamefont {Uzan}\ \emph {et~al.}(2020)\citenamefont {Uzan},
  \citenamefont {Orenstein}, \citenamefont {Jim{\'e}nez-Gal{\'a}n},
  \citenamefont {McDonald}, \citenamefont {Silva}, \citenamefont {Bruner},
  \citenamefont {Klimkin}, \citenamefont {Blanchet}, \citenamefont
  {Arusi-Parpar}, \citenamefont {Kr{\"u}ger} \emph {et~al.}}]{Uzan_2020}%
  \BibitemOpen
  \bibfield  {author} {\bibinfo {author} {\bibfnamefont {A.~J.}\ \bibnamefont
  {Uzan}}, \bibinfo {author} {\bibfnamefont {G.}~\bibnamefont {Orenstein}},
  \bibinfo {author} {\bibfnamefont {{\'A}.}~\bibnamefont
  {Jim{\'e}nez-Gal{\'a}n}}, \bibinfo {author} {\bibfnamefont {C.}~\bibnamefont
  {McDonald}}, \bibinfo {author} {\bibfnamefont {R.~E.}\ \bibnamefont {Silva}},
  \bibinfo {author} {\bibfnamefont {B.~D.}\ \bibnamefont {Bruner}}, \bibinfo
  {author} {\bibfnamefont {N.~D.}\ \bibnamefont {Klimkin}}, \bibinfo {author}
  {\bibfnamefont {V.}~\bibnamefont {Blanchet}}, \bibinfo {author}
  {\bibfnamefont {T.}~\bibnamefont {Arusi-Parpar}}, \bibinfo {author}
  {\bibfnamefont {M.}~\bibnamefont {Kr{\"u}ger}}, \emph {et~al.},\ }\bibfield
  {title} {\bibinfo {title} {Attosecond spectral singularities in solid-state
  high-harmonic generation},\ }\href@noop {} {\bibfield  {journal} {\bibinfo
  {journal} {Nature Photonics}\ }\textbf {\bibinfo {volume} {14}},\ \bibinfo
  {pages} {183} (\bibinfo {year} {2020})}\BibitemShut {NoStop}%
\bibitem [{\citenamefont {Suthar}\ \emph {et~al.}(2022)\citenamefont {Suthar},
  \citenamefont {Troj{\'a}nek}, \citenamefont {Mal{\`y}}, \citenamefont
  {Derrien},\ and\ \citenamefont {Koz{\'a}k}}]{Suthar_2022}%
  \BibitemOpen
  \bibfield  {author} {\bibinfo {author} {\bibfnamefont {P.}~\bibnamefont
  {Suthar}}, \bibinfo {author} {\bibfnamefont {F.}~\bibnamefont
  {Troj{\'a}nek}}, \bibinfo {author} {\bibfnamefont {P.}~\bibnamefont
  {Mal{\`y}}}, \bibinfo {author} {\bibfnamefont {T.~J.-Y.}\ \bibnamefont
  {Derrien}},\ and\ \bibinfo {author} {\bibfnamefont {M.}~\bibnamefont
  {Koz{\'a}k}},\ }\bibfield  {title} {\bibinfo {title} {{Role of Van Hove
  singularities and effective mass anisotropy in polarization-resolved high
  harmonic spectroscopy of silicon}},\ }\href@noop {} {\bibfield  {journal}
  {\bibinfo  {journal} {Communications Physics}\ }\textbf {\bibinfo {volume}
  {5}},\ \bibinfo {pages} {1} (\bibinfo {year} {2022})}\BibitemShut {NoStop}%
\bibitem [{\citenamefont {Vampa}\ \emph
  {et~al.}(2015{\natexlab{a}})\citenamefont {Vampa}, \citenamefont {Hammond},
  \citenamefont {Thir{\'e}}, \citenamefont {Schmidt}, \citenamefont
  {L{\'e}gar{\'e}}, \citenamefont {McDonald}, \citenamefont {Brabec},\ and\
  \citenamefont {Corkum}}]{Vampa_2015}%
  \BibitemOpen
  \bibfield  {author} {\bibinfo {author} {\bibfnamefont {G.}~\bibnamefont
  {Vampa}}, \bibinfo {author} {\bibfnamefont {T.}~\bibnamefont {Hammond}},
  \bibinfo {author} {\bibfnamefont {N.}~\bibnamefont {Thir{\'e}}}, \bibinfo
  {author} {\bibfnamefont {B.}~\bibnamefont {Schmidt}}, \bibinfo {author}
  {\bibfnamefont {F.}~\bibnamefont {L{\'e}gar{\'e}}}, \bibinfo {author}
  {\bibfnamefont {C.}~\bibnamefont {McDonald}}, \bibinfo {author}
  {\bibfnamefont {T.}~\bibnamefont {Brabec}},\ and\ \bibinfo {author}
  {\bibfnamefont {P.}~\bibnamefont {Corkum}},\ }\bibfield  {title} {\bibinfo
  {title} {Linking high harmonics from gases and solids},\ }\href@noop {}
  {\bibfield  {journal} {\bibinfo  {journal} {Nature}\ }\textbf {\bibinfo
  {volume} {522}},\ \bibinfo {pages} {462} (\bibinfo {year}
  {2015}{\natexlab{a}})}\BibitemShut {NoStop}%
\bibitem [{\citenamefont {Vampa}\ and\ \citenamefont
  {Brabec}(2017)}]{Vampa_2017}%
  \BibitemOpen
  \bibfield  {author} {\bibinfo {author} {\bibfnamefont {G.}~\bibnamefont
  {Vampa}}\ and\ \bibinfo {author} {\bibfnamefont {T.}~\bibnamefont {Brabec}},\
  }\bibfield  {title} {\bibinfo {title} {Merge of high harmonic generation from
  gases and solids and its implications for attosecond science},\ }\href@noop
  {} {\bibfield  {journal} {\bibinfo  {journal} {Journal of Physics B: Atomic,
  Molecular and Optical Physics}\ }\textbf {\bibinfo {volume} {50}},\ \bibinfo
  {pages} {083001} (\bibinfo {year} {2017})}\BibitemShut {NoStop}%
\bibitem [{\citenamefont {Gertsvolf}\ \emph {et~al.}(2008)\citenamefont
  {Gertsvolf}, \citenamefont {Jean-Ruel}, \citenamefont {Rajeev}, \citenamefont
  {Klug}, \citenamefont {Rayner},\ and\ \citenamefont
  {Corkum}}]{Gertsvolf_2008}%
  \BibitemOpen
  \bibfield  {author} {\bibinfo {author} {\bibfnamefont {M.}~\bibnamefont
  {Gertsvolf}}, \bibinfo {author} {\bibfnamefont {H.}~\bibnamefont
  {Jean-Ruel}}, \bibinfo {author} {\bibfnamefont {P.}~\bibnamefont {Rajeev}},
  \bibinfo {author} {\bibfnamefont {D.}~\bibnamefont {Klug}}, \bibinfo {author}
  {\bibfnamefont {D.}~\bibnamefont {Rayner}},\ and\ \bibinfo {author}
  {\bibfnamefont {P.}~\bibnamefont {Corkum}},\ }\bibfield  {title} {\bibinfo
  {title} {Orientation-dependent multiphoton ionization in wide band gap
  crystals},\ }\href@noop {} {\bibfield  {journal} {\bibinfo  {journal}
  {Physical Review Letters}\ }\textbf {\bibinfo {volume} {101}},\ \bibinfo
  {pages} {243001} (\bibinfo {year} {2008})}\BibitemShut {NoStop}%
\bibitem [{\citenamefont {Zhang}\ \emph {et~al.}(2021)\citenamefont {Zhang},
  \citenamefont {Huang}, \citenamefont {Li}, \citenamefont {Lan},\ and\
  \citenamefont {Lu}}]{Zhang_2021}%
  \BibitemOpen
  \bibfield  {author} {\bibinfo {author} {\bibfnamefont {Y.}~\bibnamefont
  {Zhang}}, \bibinfo {author} {\bibfnamefont {T.}~\bibnamefont {Huang}},
  \bibinfo {author} {\bibfnamefont {L.}~\bibnamefont {Li}}, \bibinfo {author}
  {\bibfnamefont {P.}~\bibnamefont {Lan}},\ and\ \bibinfo {author}
  {\bibfnamefont {P.}~\bibnamefont {Lu}},\ }\bibfield  {title} {\bibinfo
  {title} {{Intensity and wavelength dependence of anisotropic nonlinear
  absorption inside MgO}},\ }\href@noop {} {\bibfield  {journal} {\bibinfo
  {journal} {Optical and Quantum Electronics}\ }\textbf {\bibinfo {volume}
  {53}},\ \bibinfo {pages} {1} (\bibinfo {year} {2021})}\BibitemShut {NoStop}%
\bibitem [{\citenamefont {Heiner}\ \emph {et~al.}(2018)\citenamefont {Heiner},
  \citenamefont {Petrov}, \citenamefont {Steinmeyer}, \citenamefont
  {Vrakking},\ and\ \citenamefont {Mero}}]{Heiner_2018}%
  \BibitemOpen
  \bibfield  {author} {\bibinfo {author} {\bibfnamefont {Z.}~\bibnamefont
  {Heiner}}, \bibinfo {author} {\bibfnamefont {V.}~\bibnamefont {Petrov}},
  \bibinfo {author} {\bibfnamefont {G.}~\bibnamefont {Steinmeyer}}, \bibinfo
  {author} {\bibfnamefont {M.~J.}\ \bibnamefont {Vrakking}},\ and\ \bibinfo
  {author} {\bibfnamefont {M.}~\bibnamefont {Mero}},\ }\bibfield  {title}
  {\bibinfo {title} {{100-kHz, dual-beam OPA delivering high-quality, 5-cycle
  angular-dispersion-compensated mid-infrared idler pulses at 3.1 $\mu$m}},\
  }\href@noop {} {\bibfield  {journal} {\bibinfo  {journal} {Optics Express}\
  }\textbf {\bibinfo {volume} {26}},\ \bibinfo {pages} {25793} (\bibinfo {year}
  {2018})}\BibitemShut {NoStop}%
\bibitem [{\citenamefont {Gholam-Mirzaei}\ \emph {et~al.}(2017)\citenamefont
  {Gholam-Mirzaei}, \citenamefont {Beetar},\ and\ \citenamefont
  {Chini}}]{Gholam_2017}%
  \BibitemOpen
  \bibfield  {author} {\bibinfo {author} {\bibfnamefont {S.}~\bibnamefont
  {Gholam-Mirzaei}}, \bibinfo {author} {\bibfnamefont {J.}~\bibnamefont
  {Beetar}},\ and\ \bibinfo {author} {\bibfnamefont {M.}~\bibnamefont
  {Chini}},\ }\bibfield  {title} {\bibinfo {title} {{High harmonic generation
  in ZnO with a high-power mid-IR OPA}},\ }\href@noop {} {\bibfield  {journal}
  {\bibinfo  {journal} {Applied Physics Letters}\ }\textbf {\bibinfo {volume}
  {110}},\ \bibinfo {pages} {061101} (\bibinfo {year} {2017})}\BibitemShut
  {NoStop}%
\bibitem [{\citenamefont {Lindberg}\ and\ \citenamefont
  {Koch}(1988)}]{Lindberg1988}%
  \BibitemOpen
  \bibfield  {author} {\bibinfo {author} {\bibfnamefont {M.}~\bibnamefont
  {Lindberg}}\ and\ \bibinfo {author} {\bibfnamefont {S.~W.}\ \bibnamefont
  {Koch}},\ }\bibfield  {title} {\bibinfo {title} {{Effective Bloch equations
  for semiconductors}},\ }\href@noop {} {\bibfield  {journal} {\bibinfo
  {journal} {Physical Review B}\ }\textbf {\bibinfo {volume} {38}},\ \bibinfo
  {pages} {3342} (\bibinfo {year} {1988})}\BibitemShut {NoStop}%
\bibitem [{\citenamefont {Yu}\ \emph {et~al.}(2016)\citenamefont {Yu},
  \citenamefont {Zhang}, \citenamefont {Jiang}, \citenamefont {Cao},
  \citenamefont {Yuan}, \citenamefont {Wu}, \citenamefont {Bai},\ and\
  \citenamefont {Lu}}]{Yu_2016}%
  \BibitemOpen
  \bibfield  {author} {\bibinfo {author} {\bibfnamefont {C.}~\bibnamefont
  {Yu}}, \bibinfo {author} {\bibfnamefont {X.}~\bibnamefont {Zhang}}, \bibinfo
  {author} {\bibfnamefont {S.}~\bibnamefont {Jiang}}, \bibinfo {author}
  {\bibfnamefont {X.}~\bibnamefont {Cao}}, \bibinfo {author} {\bibfnamefont
  {G.}~\bibnamefont {Yuan}}, \bibinfo {author} {\bibfnamefont {T.}~\bibnamefont
  {Wu}}, \bibinfo {author} {\bibfnamefont {L.}~\bibnamefont {Bai}},\ and\
  \bibinfo {author} {\bibfnamefont {R.}~\bibnamefont {Lu}},\ }\bibfield
  {title} {\bibinfo {title} {{Dependence of high-order-harmonic generation on
  dipole moment in SiO$_2$ crystals}},\ }\href@noop {} {\bibfield  {journal}
  {\bibinfo  {journal} {Physical Review A}\ }\textbf {\bibinfo {volume} {94}},\
  \bibinfo {pages} {013846} (\bibinfo {year} {2016})}\BibitemShut {NoStop}%
\bibitem [{\citenamefont {Abbing}\ \emph {et~al.}(2022)\citenamefont {Abbing},
  \citenamefont {Campi}, \citenamefont {de~Keijzer}, \citenamefont {Morice},
  \citenamefont {Zhang}, \citenamefont {van~der Geest},\ and\ \citenamefont
  {Kraus}}]{Abbing_2022}%
  \BibitemOpen
  \bibfield  {author} {\bibinfo {author} {\bibfnamefont {S.~D.}\ \bibnamefont
  {Abbing}}, \bibinfo {author} {\bibfnamefont {F.}~\bibnamefont {Campi}},
  \bibinfo {author} {\bibfnamefont {B.}~\bibnamefont {de~Keijzer}}, \bibinfo
  {author} {\bibfnamefont {C.}~\bibnamefont {Morice}}, \bibinfo {author}
  {\bibfnamefont {Z.-Y.}\ \bibnamefont {Zhang}}, \bibinfo {author}
  {\bibfnamefont {M.~L.}\ \bibnamefont {van~der Geest}},\ and\ \bibinfo
  {author} {\bibfnamefont {P.~M.}\ \bibnamefont {Kraus}},\ }\bibfield  {title}
  {\bibinfo {title} {Efficient extreme-ultraviolet high-order wave mixing from
  laser-dressed silica},\ }\href@noop {} {\bibfield  {journal} {\bibinfo
  {journal} {arXiv preprint arXiv:2209.15561}\ } (\bibinfo {year}
  {2022})}\BibitemShut {NoStop}%
\bibitem [{\citenamefont {Vampa}\ \emph
  {et~al.}(2015{\natexlab{b}})\citenamefont {Vampa}, \citenamefont {Hammond},
  \citenamefont {Thir{\'e}}, \citenamefont {Schmidt}, \citenamefont
  {L{\'e}gar{\'e}}, \citenamefont {McDonald}, \citenamefont {Brabec},
  \citenamefont {Klug},\ and\ \citenamefont {Corkum}}]{Vampa_2015_a}%
  \BibitemOpen
  \bibfield  {author} {\bibinfo {author} {\bibfnamefont {G.}~\bibnamefont
  {Vampa}}, \bibinfo {author} {\bibfnamefont {T.}~\bibnamefont {Hammond}},
  \bibinfo {author} {\bibfnamefont {N.}~\bibnamefont {Thir{\'e}}}, \bibinfo
  {author} {\bibfnamefont {B.}~\bibnamefont {Schmidt}}, \bibinfo {author}
  {\bibfnamefont {F.}~\bibnamefont {L{\'e}gar{\'e}}}, \bibinfo {author}
  {\bibfnamefont {C.}~\bibnamefont {McDonald}}, \bibinfo {author}
  {\bibfnamefont {T.}~\bibnamefont {Brabec}}, \bibinfo {author} {\bibfnamefont
  {D.}~\bibnamefont {Klug}},\ and\ \bibinfo {author} {\bibfnamefont
  {P.}~\bibnamefont {Corkum}},\ }\bibfield  {title} {\bibinfo {title}
  {All-optical reconstruction of crystal band structure},\ }\href@noop {}
  {\bibfield  {journal} {\bibinfo  {journal} {Physical review letters}\
  }\textbf {\bibinfo {volume} {115}},\ \bibinfo {pages} {193603} (\bibinfo
  {year} {2015}{\natexlab{b}})}\BibitemShut {NoStop}%
\bibitem [{\citenamefont {Osika}\ \emph {et~al.}(2017)\citenamefont {Osika},
  \citenamefont {Chac{\'o}n}, \citenamefont {Ortmann}, \citenamefont
  {Su{\'a}rez}, \citenamefont {P{\'e}rez-Hern{\'a}ndez}, \citenamefont
  {Szafran}, \citenamefont {Ciappina}, \citenamefont {Sols}, \citenamefont
  {Landsman},\ and\ \citenamefont {Lewenstein}}]{Osika_2017}%
  \BibitemOpen
  \bibfield  {author} {\bibinfo {author} {\bibfnamefont {E.~N.}\ \bibnamefont
  {Osika}}, \bibinfo {author} {\bibfnamefont {A.}~\bibnamefont {Chac{\'o}n}},
  \bibinfo {author} {\bibfnamefont {L.}~\bibnamefont {Ortmann}}, \bibinfo
  {author} {\bibfnamefont {N.}~\bibnamefont {Su{\'a}rez}}, \bibinfo {author}
  {\bibfnamefont {J.~A.}\ \bibnamefont {P{\'e}rez-Hern{\'a}ndez}}, \bibinfo
  {author} {\bibfnamefont {B.}~\bibnamefont {Szafran}}, \bibinfo {author}
  {\bibfnamefont {M.~F.}\ \bibnamefont {Ciappina}}, \bibinfo {author}
  {\bibfnamefont {F.}~\bibnamefont {Sols}}, \bibinfo {author} {\bibfnamefont
  {A.~S.}\ \bibnamefont {Landsman}},\ and\ \bibinfo {author} {\bibfnamefont
  {M.}~\bibnamefont {Lewenstein}},\ }\bibfield  {title} {\bibinfo {title}
  {{Wannier-Bloch approach to localization in high-harmonics generation in
  solids}},\ }\href@noop {} {\bibfield  {journal} {\bibinfo  {journal}
  {Physical Review X}\ }\textbf {\bibinfo {volume} {7}},\ \bibinfo {pages}
  {021017} (\bibinfo {year} {2017})}\BibitemShut {NoStop}%
\bibitem [{\citenamefont {Parks}\ \emph {et~al.}(2020)\citenamefont {Parks},
  \citenamefont {Ernotte}, \citenamefont {Thorpe}, \citenamefont {McDonald},
  \citenamefont {Corkum}, \citenamefont {Taucer},\ and\ \citenamefont
  {Brabec}}]{Parks_2020}%
  \BibitemOpen
  \bibfield  {author} {\bibinfo {author} {\bibfnamefont {A.~M.}\ \bibnamefont
  {Parks}}, \bibinfo {author} {\bibfnamefont {G.}~\bibnamefont {Ernotte}},
  \bibinfo {author} {\bibfnamefont {A.}~\bibnamefont {Thorpe}}, \bibinfo
  {author} {\bibfnamefont {C.~R.}\ \bibnamefont {McDonald}}, \bibinfo {author}
  {\bibfnamefont {P.~B.}\ \bibnamefont {Corkum}}, \bibinfo {author}
  {\bibfnamefont {M.}~\bibnamefont {Taucer}},\ and\ \bibinfo {author}
  {\bibfnamefont {T.}~\bibnamefont {Brabec}},\ }\bibfield  {title} {\bibinfo
  {title} {Wannier quasi-classical approach to high harmonic generation in
  semiconductors},\ }\href@noop {} {\bibfield  {journal} {\bibinfo  {journal}
  {Optica}\ }\textbf {\bibinfo {volume} {7}},\ \bibinfo {pages} {1764}
  (\bibinfo {year} {2020})}\BibitemShut {NoStop}%
\bibitem [{\citenamefont {Brown}\ \emph {et~al.}(2022)\citenamefont {Brown},
  \citenamefont {Jim{\'e}nez-Gal{\'a}n}, \citenamefont {Silva},\ and\
  \citenamefont {Ivanov}}]{Brown_2022}%
  \BibitemOpen
  \bibfield  {author} {\bibinfo {author} {\bibfnamefont {G.~G.}\ \bibnamefont
  {Brown}}, \bibinfo {author} {\bibfnamefont {{\'A}.}~\bibnamefont
  {Jim{\'e}nez-Gal{\'a}n}}, \bibinfo {author} {\bibfnamefont {R.~E.}\
  \bibnamefont {Silva}},\ and\ \bibinfo {author} {\bibfnamefont
  {M.}~\bibnamefont {Ivanov}},\ }\bibfield  {title} {\bibinfo {title} {{A
  Real-Space Perspective on Dephasing in Solid-State High Harmonic
  Generation}},\ }\href@noop {} {\bibfield  {journal} {\bibinfo  {journal}
  {arXiv preprint arXiv:2210.16889}\ } (\bibinfo {year} {2022})}\BibitemShut
  {NoStop}%
\bibitem [{\citenamefont {Ambrosek}\ \emph {et~al.}(2003)\citenamefont
  {Ambrosek}, \citenamefont {Oppel}, \citenamefont {Gonz{\'a}lez},\ and\
  \citenamefont {May}}]{Ambrosek_2003}%
  \BibitemOpen
  \bibfield  {author} {\bibinfo {author} {\bibfnamefont {D.}~\bibnamefont
  {Ambrosek}}, \bibinfo {author} {\bibfnamefont {M.}~\bibnamefont {Oppel}},
  \bibinfo {author} {\bibfnamefont {L.}~\bibnamefont {Gonz{\'a}lez}},\ and\
  \bibinfo {author} {\bibfnamefont {V.}~\bibnamefont {May}},\ }\bibfield
  {title} {\bibinfo {title} {{Theory of ultrafast non-resonant multi-photon
  transitions: basics and application to CpMn (CO) 3}},\ }\href@noop {}
  {\bibfield  {journal} {\bibinfo  {journal} {Chemical Physics Letters}\
  }\textbf {\bibinfo {volume} {380}},\ \bibinfo {pages} {536} (\bibinfo {year}
  {2003})}\BibitemShut {NoStop}%
\bibitem [{\citenamefont {De~Boer}\ and\ \citenamefont
  {De~Groot}(1998)}]{DeBoer_1998}%
  \BibitemOpen
  \bibfield  {author} {\bibinfo {author} {\bibfnamefont {P.}~\bibnamefont
  {De~Boer}}\ and\ \bibinfo {author} {\bibfnamefont {R.}~\bibnamefont
  {De~Groot}},\ }\bibfield  {title} {\bibinfo {title} {{The conduction bands of
  MgO, MgS and HfO$_2$}},\ }\href@noop {} {\bibfield  {journal} {\bibinfo
  {journal} {Journal of Physics: Condensed Matter}\ }\textbf {\bibinfo {volume}
  {10}},\ \bibinfo {pages} {10241} (\bibinfo {year} {1998})}\BibitemShut
  {NoStop}%
\end{thebibliography}%

\end{document}